\documentclass[12pt,a4paper,english,aps]{revtex4-1}
\usepackage{mathptmx}
\usepackage[T1]{fontenc}
\usepackage[latin9]{inputenc}
\usepackage{verbatim}
\usepackage{units}
\usepackage{bm}
\usepackage{amsmath}
\usepackage{amssymb}
\usepackage{graphicx}
\usepackage{setspace}

\makeatletter

\newcommand{\lyxdot}{.}

\usepackage{pstricks}
\usepackage{pst-plot} 

\makeatother

\usepackage{babel}
\begin{document}

\title{Resonance line broadened quasilinear (RBQ) model for fast ion distribution
relaxation due to Alfvénic eigenmodes }

\author{N. N. Gorelenkov$^{\sharp}$, V. N. Duarte$^{\sharp\natural}$, M.
Podesta$^{\sharp}$, H. L. Berk$^{o}$ }
\email{ngorelen@pppl.gov}

\affiliation{$^{\sharp}$\emph{Princeton Plasma Physics Laboratory, Princeton
University, }}

\affiliation{$^{\natural}$\emph{Institute of Physics, University of São Paulo,
Brazil,}}

\affiliation{\emph{$^{o}$Institute for Fusion Studies, University of Texas, Austin,
Texas }}
\begin{abstract}
\noindent The burning plasma performance is limited by the confinement
of the superalfvenic fusion products such as alpha particles and the
auxiliary heating ions capable of exciting the Alfvénic eigenmodes
(AEs) \cite{GorelenkovNF14rev}. In this work the effect of AEs on
fast ions is formulated within the quasi-linear (QL) theory generalized
for this problem recently \cite{DuartePhD17}. The generalization
involves the resonance line broadened interaction of energetic particles
(EP) with AEs supplemented by the diffusion coefficients depending
on EP position in the velocity space. A new resonance broadened QL
code (or RBQ1D) based on this formulation allowing for EP diffusion
in radial direction is built and presented in details. We reduce the
wave particle interaction (WPI) dynamics to 1D case when the particle
kinetic energy is nearly constant. The diffusion equation for EP distribution
evolution is then one dimensional and is solved simultaneously for
all particles with the equation for the evolution of the wave angular
momentum. The evolution of fast ion constants of motion is governed
by the QL diffusion equations which are adapted to find the fast ion
distribution function. 

\noindent We make initial applications of the RBQ1D to DIII-D plasma
with elevated $q$-profile where the beam ions show stiff transport
properties \cite{CollinsPRL16}. AE driven fast ion profile relaxation
is studied for validations of the QL approach in realistic conditions
of beam ion driven instabilities in DIII-D. 
\end{abstract}

\maketitle

\section{Introduction}

\label{sec:Intr}Despite the significant progress in the modeling
of energetic particle (EP) driven instabiliteis in tokamaks in recent
years, we still lack the reliable quantitative and predictive capabilities
for fast ion confinement \cite{HeidbrinkNF08,GorelenkovNF14rev}.
When the dominant mechanism for fast ion transport is diffusive, a
promising reduced approach is the quasi-linear (QL) modeling which
offers the advantage of a simplified, and therefore less computationally
demanding framework \cite{Drummond_Pines_1962,VedenovSagdeev1961}.
If the mediator of EP transport is a collective instability such as
the Alfvénic eigenmode (AE) instability, the eigenmode structure and
resonances is assumed to be fixed in time. Thus the modes can be treated
perturbatively while the distribution function is allowed to evolve.
For this reason the standard QL approach does not capture the instability
frequency chirping or avalanches which are common Alfvénic spectral
response that consist of fully nonlinear oscillations. We should note
that a criterion for the likelihood of wave chirping onset (alternatively,
a criterion for the non-applicability of the QL approach) has been
recently derived and validated \cite{DuarteAxivPRL,DuartePoP2017},
which verifies the application of the QL approach for practical cases. 

While the conventional QL theory \cite{Drummond_Pines_1962,VedenovSagdeev1961,KaufmanQLPoF1972}
only applies to the situations with multiple modes overlapping (i.e.,
when the Chirikov criterion \cite{chirikov1960resonance} of resonance
overlapping is satisfied), a line broadened quasilinear model \cite{BerkNF95,BerkPoP96}
was designed to address the particle interaction with both isolated
and overlapping modes. This is done by using the same structure of
QL equations for fast ion distribution function (DF) but with the
diffusive delta function broadened along the relevant path of resonant
particles to produce a resonance line broadened quasilinear (RBQ)
diffusion. This is a key element of the RBQ model that the parametric
dependencies of the broadened window reproduce the expected saturation
levels for isolated modes. An important factor of RBQ applicability
is the presence of the velocity space diffusion \cite{MengGuoIAEA2017}
which partially includes the convective transport through the radial
dependence of the diffusion coefficient. 

The system of equations we use in the RBQ code was initially implemented
for the case of the bump-on-tail configuration in Fitzpatrick's thesis
\cite{FitzpatrickPhD97}. Ghantous \cite{GhantousPoP2014,GhantousThesis}
benchmarked this model with the Vlasov code bump-on-tail (BOT) discussing
regimes of its applicability. We revisit the RBQ model and modify
it to the realistic cases of Alfvénic instabilities excited by the
super-Alfvénic energetic ions for the first time. We express the RBQ
equations in action and angle variables, and implement them within
the NOVA/NOVA-K framework for subsequent TRANSP code simulations.
Results are presented for DIII-D discharge with the reversed shear
safety factor profile and with elevated $q_{min}$ values described
recently \cite{CollinsPRL16,HeidbrinkPOP17aecgm}. We show the predictive
capability of the RBQ model and compare it with the results of the
the kick model \cite{PodestaPPCF14,PodestaPPCF17}. 

This paper is organized as follows. The introduction is given in section
\ref{sec:Intr}. Then in section \ref{sec:RBQforml} we start the
description of the RBQ formulation by prescribing the resonance line
broadening in the Constants of Motion (COM) space to account for fast
ion dynamics near the resonances. Section \ref{sec:RBQsyst} presents
the RBQ system of differential equations written in flux and COM variables.
Section \ref{sec:WDMsec} describes the adopted Probability Density
Function (PDF) interface with the Whole Device Modeling (WDM) prototype
code TRANSP \cite{GoldstonJCP81}. Then Sec. \ref{sec:Simlt} presents
the RBQ results for selected DIII-D discharge. Finally section \ref{sec:sumsec}
summarizes the study and outlines future RBQ development. 

\section{Generalized Resonance Broadening Framework }

\label{sec:RBQforml}We introduce the QL model by describing first
the resonance line broadening which we make use of when building the
RBQ code. 

\subsection{Resonance broadening and its parametric dependencies }

\label{subsec:ResBrdng}\label{subsec:PrmtrcDep}Consider the wave
particle interaction (WPI) resonance line broadening in two-dimensional
space \cite{Dupree1966} to investigate the problem of EP transport
in the presence of realistic Alfvén eigenmodes. The QL diffusion of
a particle is along the paths of constant values of the expression
\begin{equation}
\omega P_{\varphi}+n\mathcal{E}=n\mathcal{E}^{\prime}=const,\label{eq1}
\end{equation}

where $\omega$ and $n$ are the angular frequency and toroidal mode
number of the instability. If the diffusion is approximately along
$P_{\varphi}$, i.e. at the low mode frequency the whole problem can
be reduced to the set of 1D ($\Delta\mathcal{E}\simeq0$) equations
as discussed in Fitzpatrick\textquoteright s thesis \cite{FitzpatrickPhD97}.
An appropriate rotation in space reduces the diffusion equations of
the RBQ model to the set of one-dimensional equations that are straightforward
to solve. The effect of several low-$n$ modes on energetic particles
cannot be determined without the code that captures the diffusion
in full $\mathcal{E},P_{\varphi}$ space. This is due to the complex
particle dynamics in 2D space and possible resonance overlap \cite{WhitePPCF10}. 

The conventional collisionless QL diffusion set of equations in terms
of action and angle variables can be expressed as \cite{KaufmanQLPoF1972}

\begin{equation}
\frac{\partial}{\partial t}f\left({\bf J};t\right)=\frac{\partial}{\partial{\bf J}}\left[\underline{\underline{D}}(\mathbf{J};t)\frac{\partial f}{\partial{\bf J}}\right],\label{eq:QLeq}
\end{equation}
\begin{equation}
\underline{\underline{D}}(\mathbf{J};t)=\frac{2\pi}{M^{2}}\underset{k}{\sum}\frac{C_{k}^{2}\left(t\right)}{\omega_{k}^{2}}\underset{{\bf l}}{\sum}{\bf l}{\bf l}\delta\left({\bf l}\cdot\bm{\omega_{AI}}\left({\bf J}\right)-\omega_{k}\right)\alpha_{{\bf l}}^{k}\left({\bf J}\right),\label{eq:QLeqDiffusionCoeff}
\end{equation}

\begin{equation}
\frac{dC_{k}^{2}(t)}{dt}=2\left(\gamma_{L,k}-\gamma_{d,k}\right)C_{k}^{2}(t),\label{eq:ModeAmplEvol}
\end{equation}

\begin{equation}
\gamma_{L,k}=\frac{\left(2\pi\right)^{3}}{\omega_{k}M\delta K}\int d{\bf J}\underset{{\bf l}}{\sum}\alpha_{{\bf l}}^{k}\left({\bf J}\right)\left({\bf l}\cdot\frac{\partial f}{\partial{\bf J}}\right)\pi\delta\left({\bf l}\cdot\bm{\omega_{AI}}\left({\bf J}\right)-\omega_{k}\right),\label{eq:GrowthRateKaufmann}
\end{equation}
with index $k$ denoting the mode of interest and structure contributions
(often called matrix elements) are given by 
\[
\alpha_{{\bf l}}^{k}\left({\bf J}\right)\equiv\left|\int d{\bf x}\mathbf{e}_{k}\left(\mathbf{x},\omega_{k}\right)\cdot{\bf j_{l}}^{k}\left({\bf x}\mid{\bf J}\right)\right|^{2},
\]

where $M$ is the resonant particle mass, ${\bf J}=\left(J_{1},,J_{2},,J_{3}\right)$
is the vector of actions of EP unperturbed motion (they can be substituted
by COMs in NOVA notations $\mathcal{E},P_{\varphi},\mu$), $f$ is
the EP distribution function, $\underline{\underline{D}}$ is EP diffusion
coefficient matrix. $\omega_{k}$, $C_{k}$ and $\mathbf{e}_{k}$
are the eigenmode frequency, mode amplitude and the electric field
structure. $\bm{\omega_{AI}}\equiv\left(\omega_{A1},\omega_{A2},\omega_{A3}\right)$
is the canonical (action related) frequency vector containing the
frequencies associated with the canonical angles and $\bm{l}=\left(l_{1},l_{2},l_{3}\right)$
is the vector of integer triad, $\gamma_{L,k}$ is the linear growth
of the mode, $\gamma_{d,k}$ is the wave damping rate in the absence
of the instability, $\omega_{k}\delta K$ is the mode wave energy
normalized by $C_{k}^{2}$ and ${\bf j_{l}}^{k}\left({\bf x}\mid{\bf J}\right)$
means the resonant particle current at radial point ${\bf x}$ and
having an action ${\bf J}$ (given in Appendix \ref{sec:app_gg}),
and the index $k$ relates the quantity to the $k$-th mode. The resonance
condition is represented by $\omega_{k}-{\bf l}\cdot\bm{\omega_{AI}}\left({\bf J}\right)=0$.
In the following we will use the resonance frequency notation: 
\begin{equation}
\Omega\left({\bf J}\right)\equiv{\bf l}\cdot\bm{\omega_{AI}}\left({\bf J}\right).\label{eq:gWgwal}
\end{equation}

The QL theory assumes that the mode amplitudes remain small and therefore
the theoretical coefficients are computed based on the unperturbed
orbits. In conventional QL theory, particles are considered to be
in resonance only if they exactly satisfy the wave resonance condition.
This implies that resonant particles can only diffuse over the resonant
point, which is clearly an ill-posed problem. Nonlinear effects, however,
naturally broaden the resonances. Dupree \cite{Dupree1966} realized
that the turbulent spectrum contributes to diffuse particle orbits
away from their original unperturbed trajectories. In the RBQ model
the resonant island width is incorporated into the QL theory in such
a way that it reproduces the expected saturation levels for single
modes from analytic theory \cite{BerkBreizman1990b}. The broadening
itself introduces an additional nonlinearity into the problem. The
resonance line is substituted by the broadening function $\mathcal{F}$
to replace the resonance delta function. The broadening function becomes
a more realistic platform that allows the momentum and energy exchange
between particles and waves \cite{MengGuoIAEA2017}. 

In the RBQ model the window width is determined by the sum of three
terms: 
\begin{enumerate}
\item The net growth rate ($\gamma_{k}\equiv\gamma_{L,k}+\gamma_{d,k}$,
where $\gamma_{L,k}$ and $\gamma_{d,k}$ are the linear (positive)
growth and (negative) damping rates) as expected for the wave treated
by ordinary quasilinear theory. As long as the imaginary part of the
frequency is accounted for, the diffusion coefficient naturally contains
the Lorentzian (Cauchy) distribution which has the property of having
the characteristic height of $1/\gamma_{k}$ and the full width equal
to $2\gamma_{k}$ at half maximum. The broadening is based on $\gamma_{k}$
and collapses to a delta function when $\gamma_{k}\rightarrow0$,
i.e. when the mode reaches saturation: 
\[
\pi\delta\left(\Omega\left({\bf J}\right)-\omega_{k}\right)\underset{\gamma_{k}\neq0}{\rightarrow}\frac{\gamma_{k}}{\left(\Omega\left({\bf J}\right)-Re\left\{ \omega_{k}\right\} \right)^{2}+\gamma_{k}^{2}}.
\]
\item The separatrix width expected for the wave treated by single mode
theory. In the phase space, particles that exchange energy with the
mode are trapped by the separatrix of width $4\omega_{b,k}$ \cite{MengGuoIAEA2017}.
Each particle satisfy a nonlinear pendulum equation with the given
bounce trapping frequency $\omega_{b,k}$ which leads to the phase
mixing for a single wave. 
\item The effective collisional frequency $\nu_{scatt,k}$ (as defined in
\cite{BerkPPR97,GorelenkovPoP99nlin}) since collisions imply that
particles are redistributed, being kicked in and out of the separatrix,
which leads to particle decorrelating from the resonance. This increases
the effective range of the resonance region since more particles are
allowed to interact with the mode via the resonant platform. The presence
of collisions actually simplifies the QL theory as it allows for phase
information decorrelation which gives the delta functions a broadened
width. The value of $\nu_{scatt,k}$ is sensitive to the choice of
the mode numbers and frequency. 
\end{enumerate}
It has been found \cite{FitzpatrickPhD97,GhantousPoP14} that 
\begin{equation}
\triangle\Omega_{k}\left(\mathcal{E},P_{\varphi},\mu\right)=a\omega_{b,k}+b\left|\gamma_{k}\right|\left\{ =\left|\gamma_{L,k}+\gamma_{d,k}\right|\right\} +c\nu_{scatt,k},\label{eq:Window}
\end{equation}
where the other two frequencies can be expressed as follows \cite{GorelenkovPoP99nlin}:
\begin{equation}
\nu_{scatt,k}^{3}\simeq\nu_{\perp}R^{2}\left\langle v^{2}-v_{\|}^{2}\right\rangle \left(\left.\frac{\partial\Omega}{\partial P_{\varphi}}\right|_{\mathcal{E}\text{'}}\right)^{2},\label{eq:nu3}
\end{equation}
$\nu_{\perp}$ is the $90^{0}$ pitch-angle scattering rate, $\left\langle \right\rangle $
is the drift orbit average, and 
\begin{equation}
\omega_{b,k}=\left|2C_{k}(t)V_{k}(I_{r})\left.\frac{\partial\Omega}{\partial I}\right|_{I=I_{r}}\right|^{1/2},\label{eq:gwb}
\end{equation}
where the subscript $r$ denotes the resonant location in the phase
space. The numerical constants $a$ and $c$ are chosen to best fit
the analytic theory. 

\section{RBQ System of Equations }

\label{sec:RBQsyst}For the single mode WPI case, the particle diffusion
can be projected onto the most relevant 1D path for EP dynamics in
the phase space which occurs for the constant values of the magnetic
moment $\mu$ and $\ensuremath{\mathcal{E}^{\prime}}$. Thus it is
convenient to define the following differential operator that is essentially
a gradient operator projected onto this path: 

\begin{equation}
\frac{\partial}{\partial I}=\omega_{k}\frac{\partial}{\partial\mathcal{E}}-n\frac{\partial}{\partial P_{\varphi}}=\left.\omega_{k}\frac{\partial}{\partial\mathcal{E}}\right|_{P'_{\varphi}}=\left.-n\frac{\partial}{\partial P_{\varphi}}\right|_{\mathcal{E}'}.\label{eq:ddI}
\end{equation}

Then, the 1D RBQ equation for a single mode written is 
\begin{equation}
\frac{\partial f}{\partial t}=\frac{\partial}{\partial I}\left(\sum_{k,l,m,m'}D(I;t)\right)\frac{\partial}{\partial I}f+\left(\left|\frac{\partial\Omega_{l}}{\partial I}\right|_{I_{r}}\right)^{-2}\nu_{scatt}^{3}\frac{\partial^{2}(f-f_{0})}{\partial I^{2}},\label{eq:RBQDiffOp}
\end{equation}

where 
\begin{equation}
D(I;t)=\pi n_{k}^{2}C_{k}^{2}\left(t\right)\mathcal{E}^{2}\frac{\mathcal{F}\left(I-I_{r}\right)}{\left|\frac{\partial\Omega_{l}}{\partial I}\right|}G_{m'l}^{*}G_{ml},\label{eq:Dexpr}
\end{equation}
and the matrixes $G$ are defined in Appendix \ref{sec:app_gg}. 

The growth rate is given by (cf. Appendix \ref{sec:app_gg}) 
\begin{equation}
\gamma_{L,k}=\frac{2M^{2}\pi^{3}c}{z\omega_{k}\int\rho\left|\bm{\xi}\right|^{2}d{\bf r}}\underset{\sigma_{\parallel}}{\sum}\int dP_{\varphi}d\mu d\mathcal{E}\sum_{m,m',l}G_{m'l}^{*}\mathcal{E}^{2}\tau_{b}\frac{\partial}{\partial I}fG_{ml}\frac{\mathcal{F}\left(I-I_{r}\right)}{\left|\partial\Omega_{l}/\partial I\right|},\label{eq:GammaRBQ}
\end{equation}
where $z$ is the EP electric charge, and $\tau_{b}$ is the drift
orbit period. The full system of RBQ equations is comprised by \eqref{eq:RBQDiffOp},
\eqref{eq:ModeAmplEvol}, \eqref{eq:GammaRBQ} and \eqref{eq:Window}. 

The resonances are given by (cf. Eq.\eqref{eq:gWgwal}) 

\[
\Omega_{l}\left(\mathcal{E},P_{\varphi},\mu\right)=n_{k}\left\langle \omega_{\varphi}\left(\mathcal{E},P_{\varphi},\mu\right)\right\rangle -l\left\langle \omega_{\theta}\left(\mathcal{E},P_{\varphi},\mu\right)\right\rangle =\omega_{k}
\]
where $l$ is an integer, $\omega_{\varphi}\equiv\dot{\varphi}$ and
$\omega_{\theta}\equiv\dot{\theta}$ are the toroidal and poloidal
precession frequency contributions. Note that $\Omega_{l}$ definition
of our paper is not sensitive to the properties of each $k$-th mode
but only to its toroidal mode number. 

The derivatives $\partial\Omega_{l}/\partial I$ are provided by NOVA-K.
The broadening of the resonance can be performed by choosing $\mathcal{F}$
as the\textbf{ window function} with the width $\triangle I$ that
satisfies $\int_{-\infty}^{\infty}\mathcal{F}dI=1$. The function
$\mathcal{F}$ can be arbitrarily chosen. It can, for example, be
a flat top, a Gaussian shaped or perhaps even specified via the Dupree's
window shape \cite{Dupree1966}, which for the BOT case can be transformed
to 1D window function across the resonance 
\begin{equation}
\mathcal{F}_{Dupree}=Re\int_{0}^{\infty}dt\exp\left[i\left(\Omega_{l}-\omega_{k}\right)t-D\left(\frac{\partial\Omega_{l}}{\partial I}\right)^{2}t^{3}/3\right].\label{eq:wDf}
\end{equation}
 %
{} 

\subsection{Discretized equations and boundary conditions }

Although in principle one can arbitrarily discretize the system of
equations to solve them numerically, there are some restrictions that
are implied by physical considerations. In order to conserve the system
momentum (i.e., particle plus wave momenta) at all times, the discretized
equations must guarantee internal self-consistency by adopting the
time flow presented in \cite{DuartePhD17}. This flow is chosen in
such a way that both particle diffusion and mode amplitude evolution
are calculated using the same window function at each time step. 

The linear system matrix can only be inverted (and therefore the system
can only be solved) when boundary conditions are added to the discretized
system of equations. The code needs to account for a loss boundary
in $\mathcal{E},P_{\varphi}$ plane that is different for each $\mu=const$
slice. A Neumann-type condition specifies the values that the derivative
of the distribution function takes at the boundaries of the domain.
It imposes a constant flux $\Gamma=-D\partial f/\partial I$ at the
edge. Normally in the steady state this is associated with the reflective
boundary conditions, when $D\partial f/\partial I=0$. On the other
hand, the Dirichlet-type boundary condition imposes a constraint on
the value of the function itself, such as at the loss cone when its
value is mediated by the diffusion when we set up $f=0$. 

In our diffusion solver to relax the EP distribution function, we
have the option to use either Neumann or Dirichlet boundary conditions.
They are chosen based on the following physical arguments. At the
loss boundary, $f$ should be zero and Dirichlet is physically appropriate
since it allows for particle loss. On the other hand for the inner
regions of the plasma the resonant particles are allowed to accumulate
and a Neumann condition describes the relevant physics (it is equivalent
to a reflective condition, i.e., with zero net flux). Particle number
over the plasma volume is automatically preserved if particles do
not escape at the ends. However if particles reach the ends where
they are unconfined the particle loss is quantifiable. 

\section{Probability Density Function Interface with Whole Device Modeling}

\label{sec:WDMsec}We integrate the RBQ simulations in its present
version into the transport code TRANSP which can be viewed as a prototype
of the WDM code. It enables the time-dependent integrated simulations
of a tokamak discharge. The code can be used to interpret existing
experiments as well as to develop new scenarios or make predictions
for future devices (e.g. ITER). For NB-heated discharges the NUBEAM
module within TRANSP models the evolution of the energetic particle
population based on neoclassical physics \cite{GoldstonJCP81,PankinCPC04}.
Coulomb collisions, slowing down and charge-exchange events are modeled
based on a Monte Carlo approach. To account for resonant fast ion
transport induced by Alfvénic and other MHD instabilities, NUBEAM
has been updated to include the physics-based reduced model, known
as kick model \cite{PodestaPPCF14,PodestaPPCF17}. 

For NUBEAM calculations the kick model prepares the transport probability
matrix, $p(\Delta\mathcal{E},\Delta P_{\varphi}|\mathcal{E},P_{\varphi},\mu)$,
for each instability or a set of instabilities to be included in simulations.
The matrix is defined in COM variables. For each COM grid point (or
bin) in the phase space, $p(\Delta\mathcal{E},\Delta P_{\varphi})$
represents the probability of $\mathcal{E}$ and $P_{\varphi}$ changes
(kicks) experienced by the energetic ions as a result of their interaction
with the instability. 

In previous works \cite{PodestaNF16,PodestaPoP16,PodestaPPCF17} the
transport probability matrices were computed numerically by the guiding
center code ORBIT \cite{WhitePoP84} using the mode structures from
the NOVA code (details on how the matrix is computed for a given instability
are found in the Appendix of Ref.\cite{PodestaPPCF17}). In this work
the quasi-linear diffusion coefficient computed by the RBQ1D code
is used instead to reconstruct the $p(\Delta\mathcal{E},\Delta P_{\varphi}|\mathcal{E},P_{\varphi},\mu)$
probabilities under the assumption of negligible energy variations
induced by the wave-particle interactions. 

Consider the trajectory of a resonant particle subject to constraints
in the $(\mathcal{E},P_{\varphi},\mu)$ space \cite{WhitePPCF11,WhiteCSNS12}:
Equation \eqref{eq1} implies that for a single mode the variations
in $\mathcal{E}$ and $P_{\varphi}$ are related through 
\begin{equation}
\Delta P_{\varphi}/\Delta\mathcal{E}=n/\omega.\label{eq2}
\end{equation}

Based on this constraint and under the assumption of diffusive transport
(implied by the RBQ1D formulation), the bi-variate PDF for $\Delta\mathcal{E}$
and $\Delta P_{\varphi}$ changes can be represented as: 

\begin{equation}
p\left(\Delta\mathcal{E},\Delta P_{\varphi}|P_{\varphi},\mathcal{E},\mu,A_{kick}\right)=p_{0}e^{-\left[\frac{\left(\Delta\mathcal{E}-\Delta\mathcal{E}_{0}\right)^{2}}{\sigma_{\mathcal{E}}^{2}}+\frac{\left(\Delta P_{\varphi}-\Delta P_{\varphi0}\right)^{2}}{\sigma_{P_{\varphi}}^{2}}-2\rho\frac{(\Delta\mathcal{E}-\Delta\mathcal{E}_{0})(\Delta P_{\varphi}-\Delta P_{\varphi0})}{\sigma_{\mathcal{E}}\sigma_{P_{\varphi}}}\right]/2\left(1-\rho\right)},\label{eq3}
\end{equation}

with the normalization factor 
\begin{equation}
p_{0}=\frac{1}{2\pi\sigma_{\mathcal{E}}\sigma_{P_{\varphi}}\sqrt{1-\rho^{2}}}.\label{eq4}
\end{equation}

The correlation parameter $\rho=\frac{\left\langle \left(\Delta\mathcal{E}-\Delta\mathcal{E}_{0}\right)\left(\Delta P_{\varphi}-\Delta P_{\varphi0}\right)\right\rangle }{\sigma_{\mathcal{E}}\,\sigma_{P_{\varphi}}}$
takes into account coupling between $\Delta\mathcal{E}$ and $\Delta P_{\varphi}$
expressed in Eqs.(\ref{eq1},\ref{eq2}), where angular brackets mean
energy and $P_{\varphi}$ averaging weighted by PDF. Note that the
offset (or convective) terms $\Delta\mathcal{E}_{0}$ and $\Delta P_{\varphi0}$
are vanishing for the cases when there is no systematic drift in energy
or $P_{\varphi}$. The variances $\sigma_{\mathcal{E}}$ and $\sigma_{P_{\varphi}}$
are related to the diffusion coefficients in energy and canonical
angular momentum, $D_{\mathcal{E}}$ and $D_{P_{\varphi}}$, and give
the spread of the distribution along the $\Delta\mathcal{E}$ and
$\Delta P_{\varphi}$ axes: 

\begin{equation}
\sigma_{\mathcal{E}}^{2}=4D_{\mathcal{E}}\delta t~~;~~\sigma_{P_{\varphi}}^{2}=4D_{P_{\varphi}}\delta t\label{eq5}
\end{equation}
which allows the computation of the transport probability matrix over
the time step $\delta t$ associated with resonant particles for known
quasi-linear diffusivities from the RBQ1D model. 

In general, the probability matrix from Eqs.(\ref{eq3}-\ref{eq5})
corresponds to the WPI contribution to the total transport probability.
Another term accounts for the fact that not all particles in a specific
$\left(\mathcal{E},P_{\varphi},\mu\right)$ bin are necessarily resonant
with a given instability. In fact, there can be large portions of
phase space where no resonances are present. In general, for each
discrete $\left(\mathcal{E},P_{\varphi},\mu\right)$ bin there is
a fraction $k_{res}$ of resonant particles and a fraction $k_{non-res}$
of unperturbed, non-resonant particles that are not affected by the
instability. To compute those fractions one can infer the volume $V_{res}$
occupied by resonant ions within the $\left(\mathcal{E},P_{\varphi},\mu\right)$
bin and compute the ratio with respect to the total bin volume $V_{tot}$:
\begin{equation}
k_{res}=\frac{V_{res}}{V_{tot}}~~;~~k_{non-res}=1-\frac{V_{res}}{V_{tot}}.\label{eq6}
\end{equation}
Once these two fractions are known for each bin the total probability
is 
\[
p\left(\Delta\mathcal{E},\Delta P_{\varphi}\right)=k_{non-res}\delta\left(\mathcal{E}=0,P_{\varphi}=0\right)+k_{res}p_{res}\left(\Delta\mathcal{E},\Delta P_{\varphi}|\mathcal{E},P_{\varphi},\mu\right)
\]
with $\delta(\mathcal{E}=0,P_{\varphi}=0)$ the delta function and
$p_{res}$ from Eq.\eqref{eq3}. 

We have outlined above the computations of PDF matrices for fast ion
diffusion in COM space for subsequent NUBEAM calculations. 

\section{Applications to Critical Gradient DIII-D Experiments }

\label{sec:Simlt}
\begin{figure}
\begin{singlespace}
\begin{centering}
\includegraphics[bb=0bp 0bp 350bp 220bp,clip,scale=0.7]{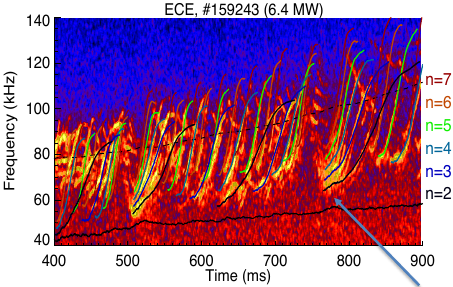}\hspace{-200pt}\raisebox{116pt}{\textcolor{yellow}{\normalsize (a)}}\hspace{190pt}\hspace{-60pt}\raisebox{-15pt}{\normalsize time of interest}\hspace{-15pt}\includegraphics[bb=0bp 0bp 450bp 350bp,clip,scale=0.45]{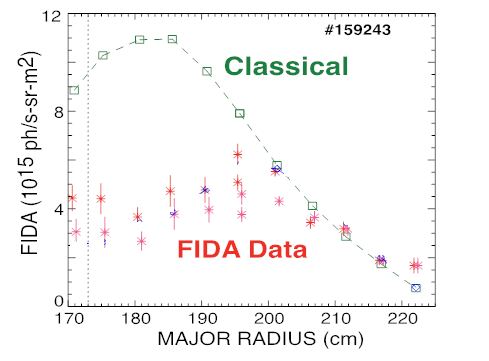}\hspace{-45pt}\raisebox{116pt}{\normalsize (b)}\hspace{43pt}
\par\end{centering}
\end{singlespace}
\caption{\emph{A $CO_{2}$ interferometer power spectra for DIII-D shot $\#159243$
during the current ramp with $6.5MW$ NBI. The spectrum on figure
(a) shows multiple sweeping fre quency Reversed Share AEs (RSAE) and
steady Toroidicity-induced AE (TAE) modes. Figure (b) compares classical
TRANSP predictions for beam ion profile. The graph shows the error
bar uncertainty associated with the background subtraction. The two
sets of data represent two different light calibrations. The dotted
vertical line indicates the location of the magnetic axis. (Figures
are reproduced from Ref.}\cite{HeidbrinkPOP17aecgm}\emph{) \label{fig:Mirnov} }}
\end{figure}

We are chosing recent DIII-D experimental studies
for RBQ application with the goal of performing initial validations.
In those experiments several ubiquitous EP transport responses were
recognized: (i) EP transport suddently changes at bifurcation; (ii)
the transport is intermittent in time; (iii) EP profiles are resilient
to the changes in beam injections \cite{CollinsPRL16}. The AE power
spectrum increases linearly with the total driving beam power above
EP threshold level. One representative discharge is chosen, $\#159243$,
with $6.4MW$ of tangential beam power for our analysis. One time
slice is studied extensively, $t=805msec$, when the reversed shear
magnetic safety factor profile minimum value turns lower than $q_{min}=3$.
Its spectrogram is depicted in Fig.\eqref{fig:Mirnov} where the point
of interest is indicated by the arrow. 

An interesting feature of the data was revealed later when the velocity
space resolution allowed the demonstration of a rather unexpected
hollow profile of EP distribution function. The data was collected
in part by the Fast Ion D$_{\alpha}$ (FIDA) diagnostics with some
velocity space resolution. A similar feature was found using the interpretive
kick model where the integration of the EP distribution function was
implemented along the same FIDA ``view'' window COM path. An application
of the critical gradient model \cite{GorelenkovNF16cgm} to the same
case did not find any hollow EP profile behaviour but rather monotonic
radial EP profiles. 

\subsection{Perturbative NOVA simulations for RBQ}

Since RBQ is essentially perturbative and works as a postprocessor
for NOVA/NOVA-K runs, its analysis is initiated by identifying the
mode structures of AE instabilities responsible for EP transport.
Extensive efforts were already undertaken with the kick model applications
recently \cite{HeidbrinkPOP17aecgm}. We make use of these RSAE/TAE
results. 

\begin{figure}
\begin{singlespace}
\begin{centering}
\includegraphics[bb=0bp 0bp 550bp 336bp,clip,scale=0.5]{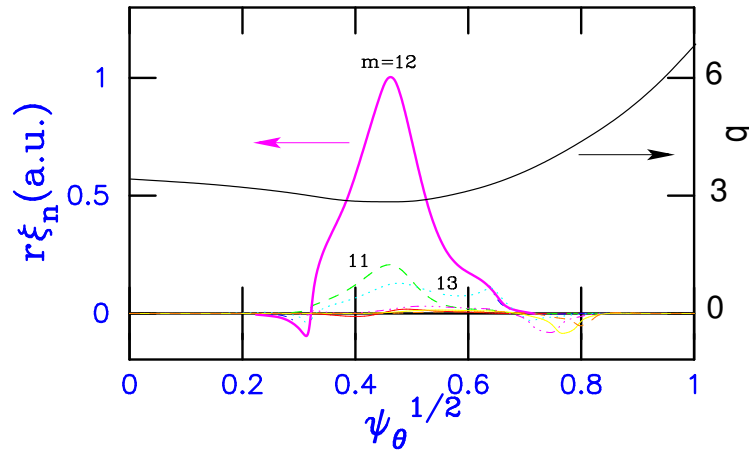}
\par\end{centering}
\end{singlespace}
\begin{singlespace}
\caption{\emph{A plot of the $q$-profile of the plasma and spatial structure
of poloidal harmonics for the radial displacement $n=4$, $f=84kHz$
RSAE computed by the ideal MHD code NOVA. The mode structure is computed
for the plasma equilibrium with the shown safety factor profile. \label{fig:n4RSAE}}}
\end{singlespace}
\end{figure}

To show the details of RBQ analysis we choose a $n=4$ RSAE from Ref.\cite{HeidbrinkPOP17aecgm}
for EP distribution relaxation. The mode is localized near $q_{min}\simeq3$
at $t=805msec$ surface and has one dominant $m=12$ poloidal harmonic
which is shown in Fig. \ref{fig:n4RSAE}. 

The ideal MHD NOVA mode structures is used by the NOVA-K code \cite{GorelenkovPoP99}
to evaluate the wave particle interaction (WPI) matrices for further
processing by the RBQ code as described in Sec.\ref{sec:RBQforml}
and the Appendix \ref{sec:app_gg}. For the RSAE shown in Fig. \ref{fig:n4RSAE},
NOVA-K computes the normalized growth rate $\gamma_{L}/\omega=3.2\%$
and the total damping rate $\gamma_{d}/\omega=-1.81\%$. 

\begin{figure}
\begin{centering}
\vspace*{-2.9cm}
\par\end{centering}
\begin{singlespace}
\begin{centering}
\includegraphics[bb=0bp 5bp 530bp 536bp,clip,scale=0.4]{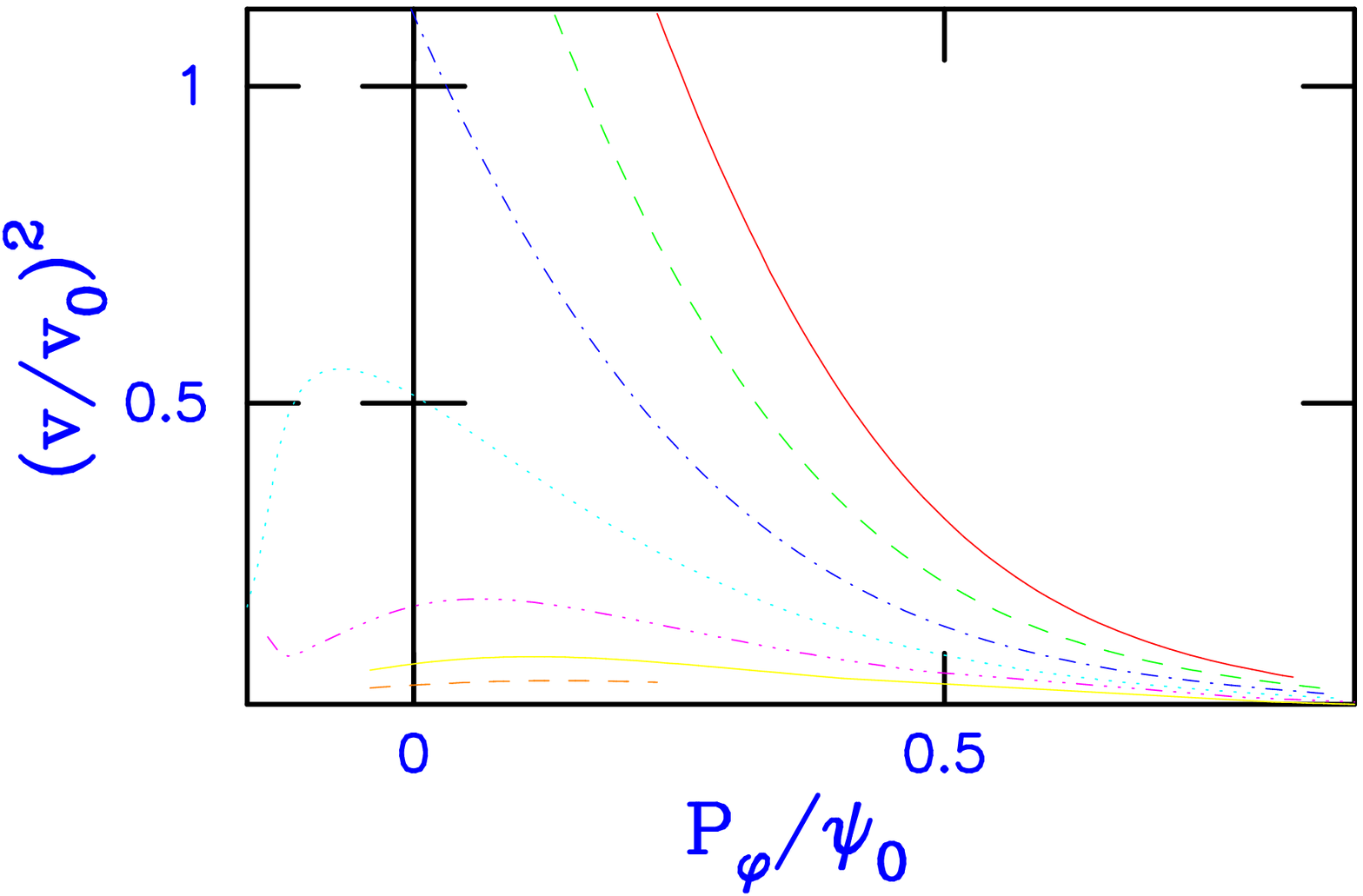}
\hspace{-40pt}\raisebox{116pt}{(a)}\hspace{30pt}\includegraphics[bb=40bp 60bp 650bp 380bp,clip,width=7cm,height=4.7cm]{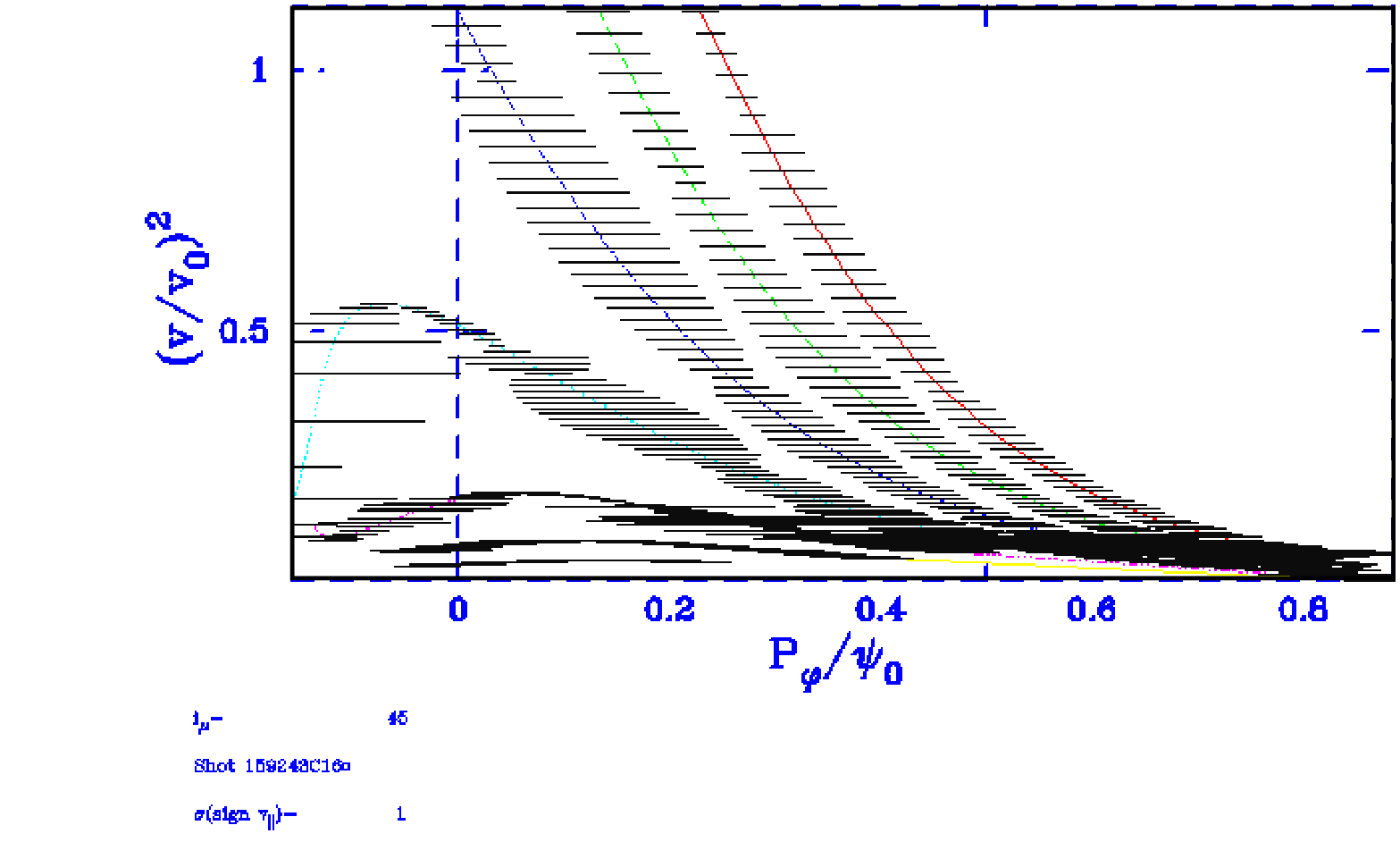}\hspace{-35pt}\raisebox{116pt}{\normalsize (b)}\hspace{45pt}
\par\end{centering}
\end{singlespace}
\caption{\emph{Resonance lines of passing EPs interacting with the RSAE mode
(the mode structure is shown in Fig.\ref{fig:n4RSAE}) in the COM
plane: normalized kinetic energy vs. canonical toroidal momentum.
Shown in figure (a) are seven dominant resonancies at }$\lambda=0.4$\emph{.
Figure (b) represents the broadening of the resonance lines in the
direction of $P_{\varphi}$ computed at the mode amplitude $\delta B/B=7.4\times10^{-3}$.
\label{fig:n4RSAEresbrdn} }}
\end{figure}

As a result the broadening of the resonance line, Eq.\eqref{eq:Window}
is computed at each resonance point within the RBQ code as illustrated
in Fig.\ref{fig:n4RSAEresbrdn}. Figure \ref{fig:n4RSAEresbrdn} (a)
shows 7 resonance lines at one value of $\lambda=0.4$ corresponding
to co-passing ions, where $\lambda=\mu B_{0}/\mathcal{E}$ and $B_{0}$
is the magnetic field strength on the axis. Figure \ref{fig:n4RSAEresbrdn}
(b) shows the broadening of those resonances due to the first and
the last terms in Eq.\eqref{eq:Window}. 

The relaxation of EP distribution function (DF) can be computed within
the RBQ simulations accurately. However in the present form the RBQ1D
code computes the WPI induced diffusion into the TRANSP code through
the probability density function (PDF) introduced above (see Sec.\ref{sec:WDMsec}
and also Ref. \cite{PodestaNF15}). At the moment we will limit the
initial RBQ1D application to such interface with TRANSP. 

It has been shown that the kick model captures the EP diffusion in
the velocity space which is substantially different from the diffusion
``ad-hoc'' model used normally in TRANSP and is a significant factor
affecting the EP distribution function \cite{PodestaNF16}. We employ
the AE driven PDFs computed by RBQ within TRANSP and show the obtained
results in the next figure \ref{fig:n4RSAEDF}. The results are for
the single RSAE shown in Fig. \ref{fig:n4RSAE}. 
\begin{figure}
\begin{centering}
\hspace{120pt}\raisebox{82pt}{\normalsize (a)}\hspace{-140pt}\includegraphics[bb=40bp 25bp 500bp 160bp,clip,width=0.5\textwidth,height=0.15\textheight]{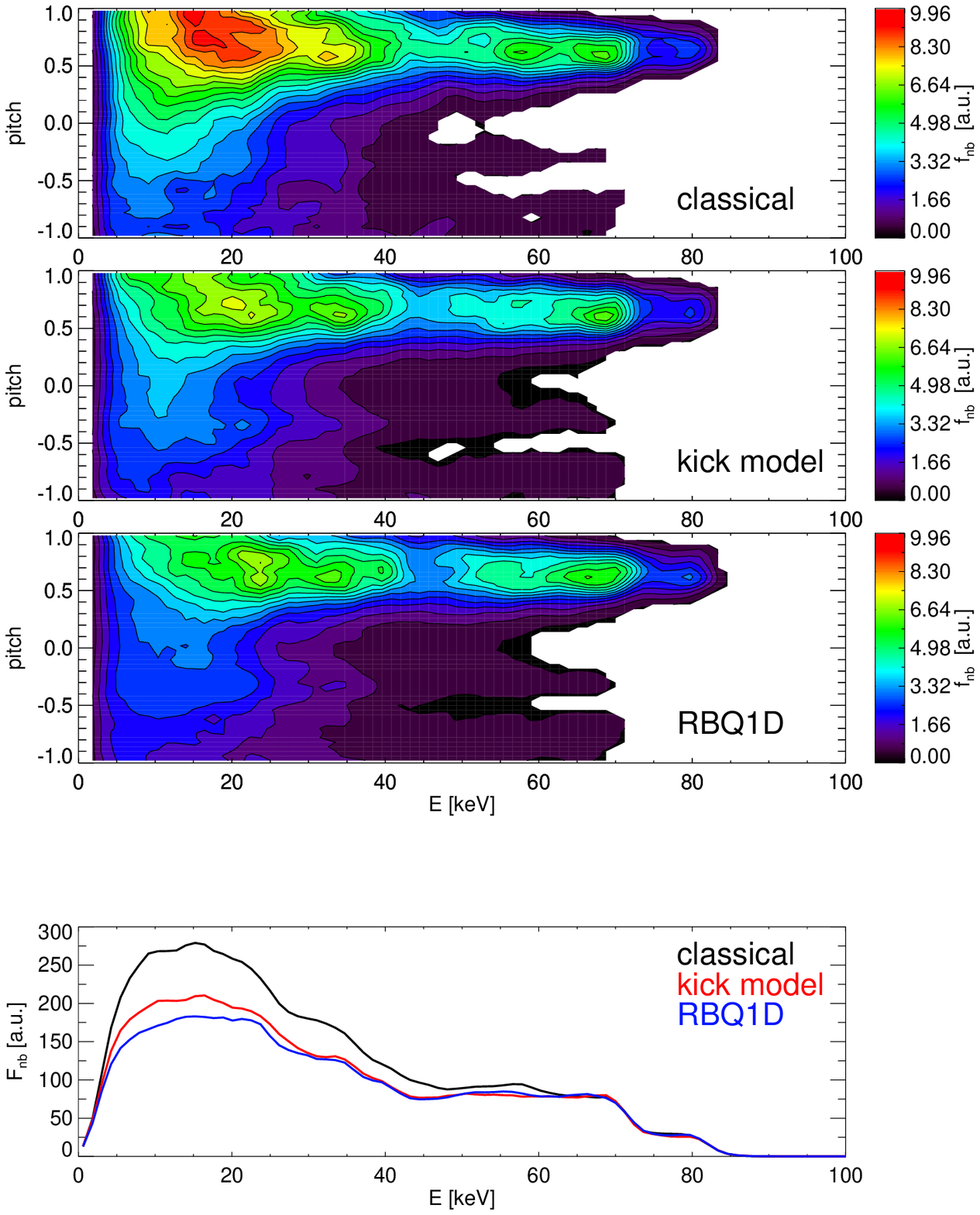}\hspace{130pt} \raisebox{170pt}{\normalsize (b)} \hspace{-19pt} \raisebox{115pt}{\normalsize (c)} \hspace{-19pt} \raisebox{60pt}{\normalsize (d)} \hspace{-200pt}\includegraphics[bb=0bp 180bp 450bp 560bp,clip,scale=0.5]{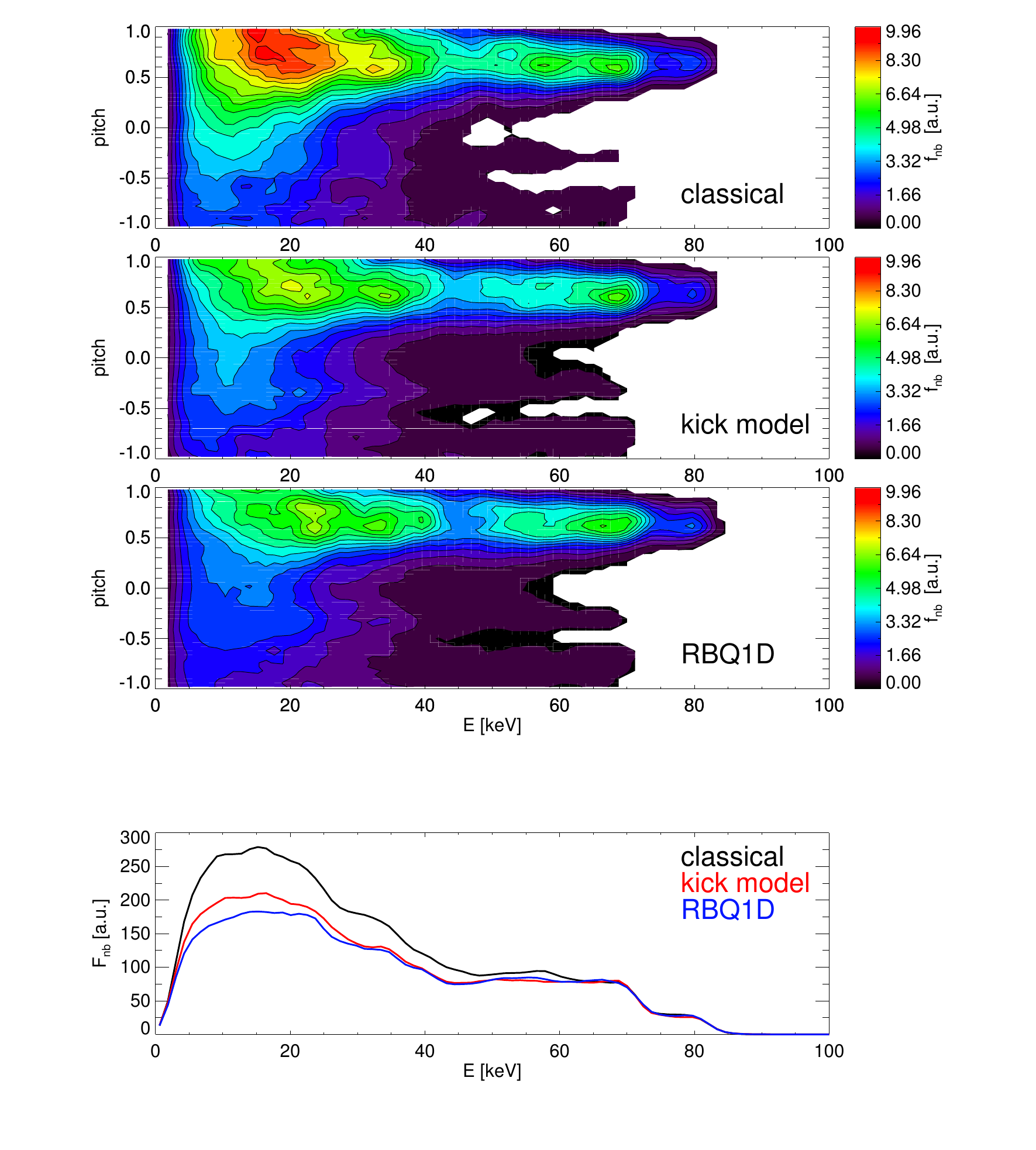}\vspace{-10pt}
\par\end{centering}
\begin{singlespace}
\caption{\emph{TRANSP simulated EP distribution function in the velocity space
for the time of interest, $t=800msec$ using the diffusion computed
by RBQ1D and the kick model as indicated. Shown are DF averaged between
the normalized poloidal flux $\sqrt{\psi/\psi_{w}}=0.4$ and $0.6$
values, where $\psi_{w}$ is the poloidal magnetic field flux at the
wall. Figure (a) shows the pitch angle averaged DF velocity dependance.
Figures (b,c,d) correspond to the classical, kick model and RBQ1D
simulations of the EP diffusion. \label{fig:n4RSAEDF}}}
\end{singlespace}
\end{figure}

Comparing three cases in the Fig.\ref{fig:n4RSAEDF} tells us that
the effect of AEs is similar on the beam ions if modeled by RBQ1D
or by the kick models. Insert (a) tells us that most of the losses
are at the low energies, $E_{b}=10-30keV$, where neutrons are generated.
This is consistent with the Fig. \ref{fig:n4RSAEresbrdn}(b) which
shows that this is where the resonances are most overlapped which
results in stronger radial transport. 

RBQ has been implemented in a way which allows either interpretive
or predictive analysis. Let us consider them both. 

\subsection{Interpretive RBQ implementation for TRANSP code}

\label{subsec:Inrpt}The present mode of RBQ operation relies on AE
amplitude values inferred from the experimental measurements. As we
mentioned above, the kick model has been successful describing DIII-D
FIDA experiments by computing EP diffusion specific to particle position
in the COM space \cite{HeidbrinkPOP17aecgm}. It reproduced the hollow
EP pressure profiles as a result. We use the same amplitude values
for RBQ1D analysis. 

\begin{figure}
\begin{singlespace}
\begin{centering}
\hspace{-150pt}{\small{}\includegraphics[bb=0bp 165bp 220bp 318bp,clip,scale=1.085]{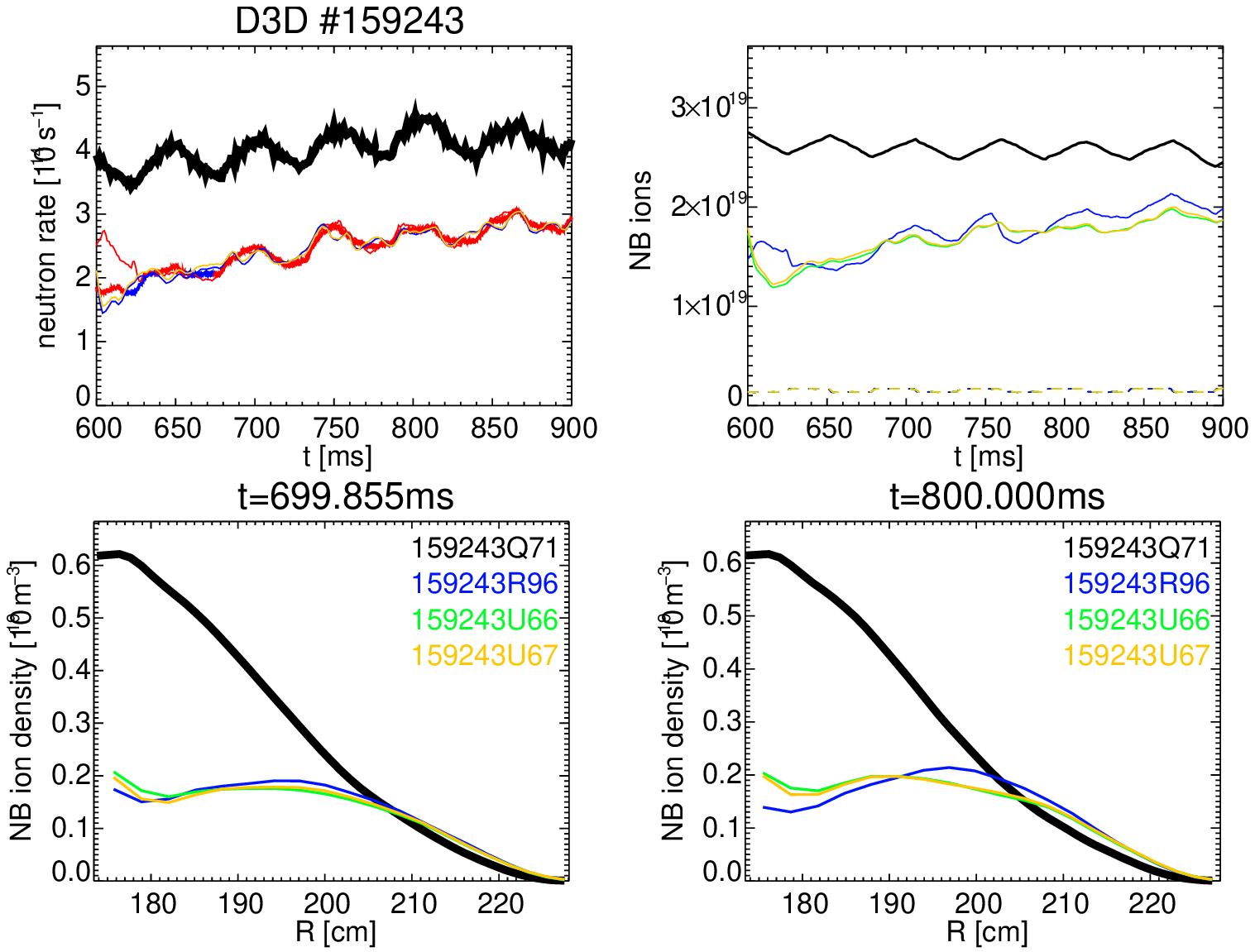}}\hspace{-175pt}\raisebox{150pt}{(a)}\hspace{165pt}{\small{}\includegraphics[bb=0bp 1bp 440bp 233bp,clip,scale=0.72]{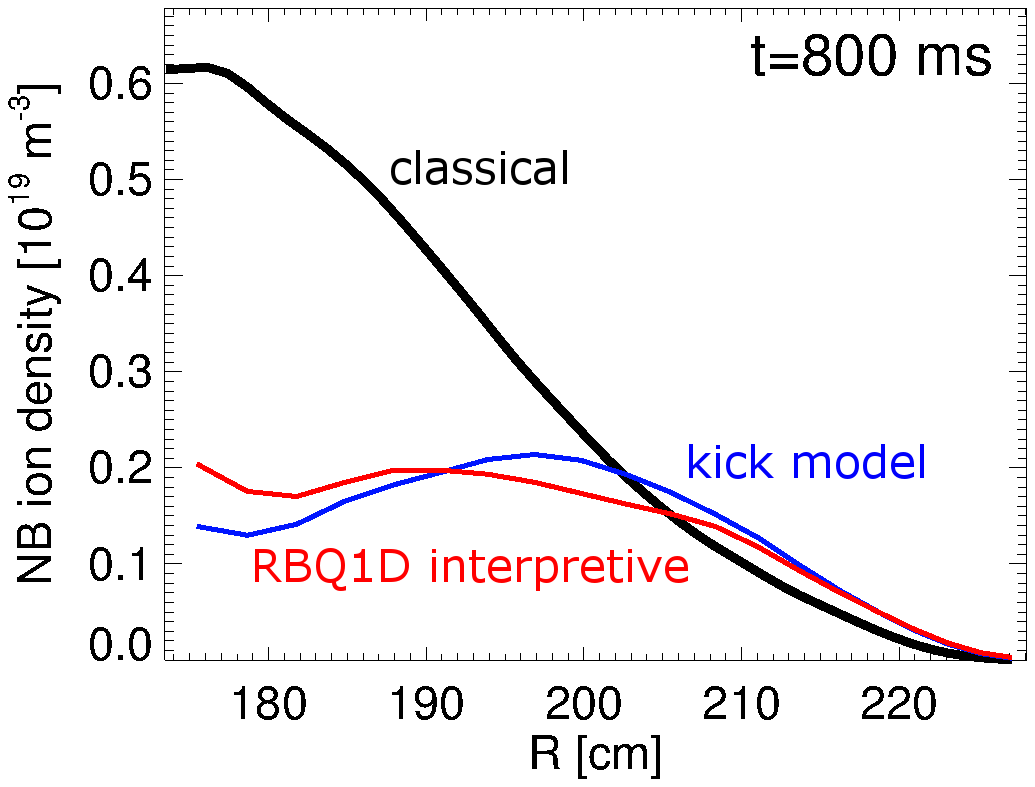}}\hspace{-260pt}\raisebox{150pt}{\normalsize (b)}\hspace{20pt}
\par\end{centering}
\end{singlespace}
\caption{\emph{Left figure compares the neutron rate computed by TRANSP using
various assumptions about EP diffusivities. Top (black, classical)
dependence is obtained ignoring EP diffusion. Kick model NBI density
profiles are blue curves whereas RBQ1D profile and time dependencies
are shown in red.} \label{fig:Intrprprfls} }
\end{figure}

Details of TRANSP computations using PDFs are already published \cite{PodestaNF16,PodestaPoP16,PodestaPPCF17},
so that here we present the results using those techniques. We add
the RBQ prescribed PDFs and summarize this exercise in figure \ref{fig:Intrprprfls}.
One can see that both models work well by computing beam ion hollow
density profils. We should add however that in the inerpretive mode
RBQ1D and kick models had additional constraint due to the neutron
flux. 

Additional studies were done which show the origin of the inverse
profile behaviour. We have found that the origin of the reversed profiles
is due to the diffusion of copassing beam ions subject to strong diffusion.
They are dominating the EP population near the center and are preferentially
transported in radial direction outward. 

\subsection{Predictive RBQ implementation for TRANSP code}

\label{subsec:prdctv}In the predictive RBQ analysis we repeated computations
from the previous section \ref{subsec:Inrpt} by using the same set
of AEs. However we compute AE amplitudes in a different manner by
finding their values when the particular mode reaches the saturation,
i.e. when the growth rate balances the damping rate, $\gamma_{L,k}=-\gamma_{d,k}$.
This way we are not addressing the intermittency in AE transport seen
in experiments (see Fig.\ref{fig:Mirnov}), nor do we address the
long time scale changes in AE stability as the RSAEs, for example,
are sweeping its frequency on a hundred milisecond time scale. 

\begin{figure}
\begin{centering}
\hspace{-100pt}\includegraphics[bb=0bp 0bp 580bp 430bp,clip,scale=0.32]{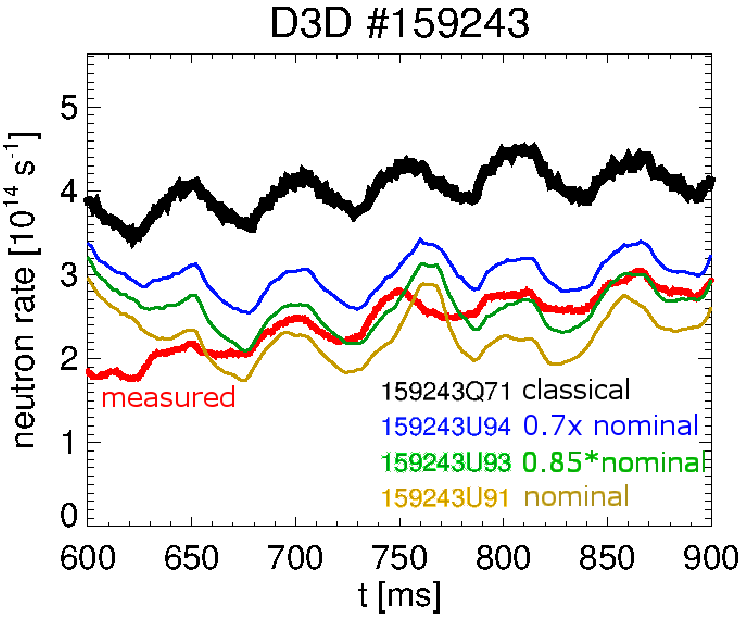}\hspace{-145pt}\raisebox{120pt}{(a)}\hspace{165pt}\includegraphics[bb=0bp 0bp 580bp 430bp,clip,scale=0.32]{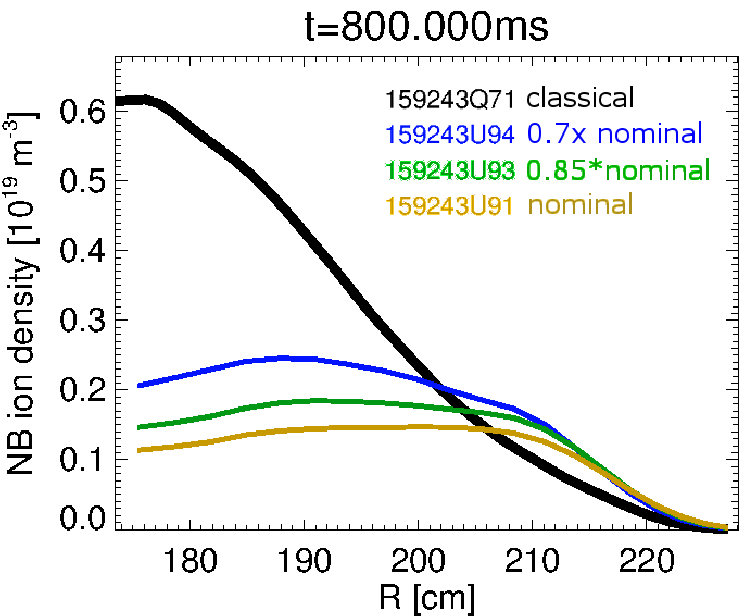}\hspace{-120pt}\raisebox{120pt}{\normalsize (b)}\hspace{20pt}
\par\end{centering}
\caption{\emph{The same as in Fig. \ref{fig:Intrprprfls} but computed within
the predictive TRANSP analysis. Different EP profile and evolution
curves are color coded as indicated on the figure. \label{fig:prdctvprfls} }}
\end{figure}

The application of RBQ predictions within the TRANSP simulations are
summarized in Figs. \ref{fig:prdctvprfls}. After TRANSP turns AE
diffusion on at $t=600ms$ the diffusion coefficients are kept constant
throughout simulations. Variations in AE amplitudes are reflected
by different TRANSP computations. We are showing that the predictive
RBQ runs with 0.85 times the amplitude values which fit the best the
neutron defiict in TRANSP simulations. The beam density profiles remains
hollow within the amplitude variations. 

\section{Summary and future plans}

\label{sec:sumsec}This paper demonstated the effectiveness of the
quasi-linear model in its applications for realistic simulations of
the beam ion self-induced relaxation via the Alfvénic instabilities.
We have summarized the formulation for the Resonance line Broadened
QL (RBQ) numerical model with the EP diffusion near the resonances.
The formulation can be applied for isolated or for the overlapping
modes. 

The RBQ1D has been applied in the interpretive and in the predictive
mode to DIII-D critical gradient experiments via prescription of the
PDF for the beam ions. Initial results show that the model is sound
and ready to be applied to predict the fast ion relaxation in burning
plasmas. However more validations are required in order to gain the
confidence in its predictive capability. 

Among the immediate plans we have the development of the RBQ in its
2D version where the EP diffusion paths are sensitive to AE toroidal
mode number and its frequency. The number of refined grid points of
energetic particle diffusion and the need to resolve the resonances
make the 2D problem intrinsically complex and computationally demanding. 

\section*{Acknowledgement}

This manuscript has been authored by Princeton University under Contract
Number DE-AC02-09CH11466 and by DIII-D National Fusion Facility under
Contract Number DE-FG02- 95ER54309 with the U.S. Department of Energy
and by the São Paulo Research Foundation (FAPESP, Brazil) under grants
2012/22830-2 and 2014/03289-4. The United States Government retains
and the publisher, by accepting the article for publication, acknowledges
that the United States Government retains a non-exclusive, paid-up,
irrevocable, world-wide license to publish or reproduce the published
form of this manuscript, or allow others to do so, for United States
Government purposes. 

We thank Dr. R. Nazikian for motivating valuable discussions. 

\appendix

\section{Linear growth rate as a result of the nonadiabatic component of the
distribution function }

\label{sec:app_gg}NOVA \cite{ChengAP85,GorelenkovPoP99} is a nonvariational,
ideal MHD code primarily used to integrate non-Hermitian %
{} eigenmode equations in the presence of EPs, using a general flux
coordinate system. NOVA-K \cite{ChengPhR92} is a stability code used
to study the destabilization of TAEs by EPs free energy stored in
the gradients of the distribution. The resonance response of energetic
particles enter the system through the perturbed pressure associated
with them. NOVA makes no use of inverse aspect ratio approximation
and hence is well suited to study spherical tokamaks. The code uses
Fourier expansion in $\theta$ and cubic spline finite elements in
the radial $\psi$ direction. Following the procedure of Cheng \cite{ChengPhR92}
we may start with the MHD formalism where the fast particle contribution
is treated perturbatively, with all perturbed quantities represented
in the form

\[
A\left(\bm{r},t\right)=\underset{m}{\sum}A_{m}e^{i\left(S_{m}-\omega t\right)};\quad S_{m}\equiv m\theta-n\varphi
\]
We shall start with the momentum conservation equation. This equation
can be put in a quadratic form if dot multiplied by $\bm{\xi}^{*}$
and integrated over the whole plasma volume. If eqs. (3.15) of \cite{GorelenkovPoP99}
are used we get

\[
D(\omega)=\delta W_{f}+\delta W_{k}-\delta K=0
\]
where the inertial (kinetic) energy is given by $\delta K=\omega^{2}\int\rho\left|\bm{\xi}\right|^{2}d{\bf r}$.
$\delta W_{f}$ is the total fluid potential energy and $\delta W_{k}$
is the EPs potential energy .\footnote{In NOVA-K, both potential and kinetic energies are defined as twice
their actual values. This can be seen from comparing equations (3.69)
and (3.70) of \cite{ChengPhR92} with equations (4.19) and (4.31)
of \cite{WhiteTH13}. This choice does not introduce changes to the
growth rate.} $\omega_{r}\delta W_{k}$ is the power being released by the resonant
particles. The quadratic form is particularly useful when stability
analysis is addressed. For example, if the mode frequency is assumed
to be $\omega=\omega_{r}+i\gamma$, with $\left|\gamma\right|\ll\left|\omega_{r}\right|$,
it is obtained \cite{ChengPhR92}

\[
\gamma\approx\frac{\Im\delta W_{k}}{2\delta K}\omega_{r}
\]
In the last equation, it was used that the inertial energy is close
to the mode energy. This is because Alfvén waves involve negligible
electric field perturbations. Their energy is nearly equally divided
between the perturbed magnetic energy and the kinetic energy of particles
oscillating as a result of the perturbation.$\Re$ and $\Im$ denote
the real and imaginary parts, respectively. $\delta K=\omega^{2}\int\rho\left|\bm{\xi}\right|^{2}d{\bf r}$
accounts for all poloidal harmonics and is simply a number, being
a global factor for each toroidal mode number $n$. Since $\rho$
is the total plasma density, it is very little affected by the fast
ions density. This number is also provided by NOVA-K. For TAEs, the
growth rate is not simple as in the case of an idealized bump-on-tail
configuration, being an integral over the resonant curve in phase
space and depending on the mode structure. In order to compute $\Im\delta W_{k}$,
and consequently $\gamma_{n}$, the non-adiabatic part of the distribution
function, $g$, must be calculated. This function can be redefined
as $\hat{g}$ to satisfy 
\[
\frac{d\hat{g}}{dt}=\frac{z}{M}\frac{\partial f}{\partial\mathcal{E}}\left(\omega-\omega_{*}\right)\hat{X}
\]
where $\omega_{*}=n\left(\partial f/\partial P_{\varphi}\right)/\left(\partial f/\partial\mathcal{E}\right)$
is the diamagnetic frequency, being a measure of the relation of the
radial gradient in the energetic in EP profiles to the velocity gradient.
Since $dP_{\varphi}=-\frac{qr}{\omega_{c}}dr$ and the EP density
decreases with radius, we have $\frac{\partial f}{\partial P_{\varphi}}>0$.
On the other hand, the average energy of EPs also decreases with radius,
which implies $\frac{\partial f}{\partial\mathcal{E}}<0$. Therefore,
in a tokamak, the free energy stored in the radial gradient drives
the mode while the negative gradient in energy tends to stabilize
the mode. The mode is unstable when $\omega_{*}>\omega$. $\hat{X}$
is given by
\[
\hat{X}=\left(\frac{d{\bf r}_{c}}{dt}-v_{\parallel}\hat{b}\right)J_{0}\nabla\phi_{c}+\frac{i\mu\omega M}{z}\frac{df}{d\mu}\left(J_{0}+J_{2}\right)B_{\parallel c}
\]
where the subscript $c$ means that the quantity is evaluated at the
particle gyrocenter. The Bessel functions are understood to have the
argument $v_{\perp}\nabla_{\perp}/\omega_{c}$, which operates on
the perturbed quantities. $v_{\perp}\nabla_{\perp}/\omega_{c}\sim v_{\perp}k_{\perp}/\omega_{c}\ll1$
which justifies the Taylor expansion of the Bessel functions, $J_{\alpha}\left(v_{\perp}\nabla_{\perp}/\omega_{c}\right)\rightarrow\overset{\infty}{\underset{j=0}{\sum}}\frac{\left(-1\right)^{j}}{j!\Gamma\left(j+\alpha+1\right)}\left(v_{\perp}\nabla_{\perp}/\omega_{c}\right)^{2j+\alpha}$.
If ${\bf E}_{\perp}=-\nabla\phi$ (${\bf A}_{\perp}$ contribution
neglected) we can write 

\[
\hat{X}\simeq\frac{i\omega M}{z}\left[\left(2\mathcal{E}-3\mu B\right)J_{0}\bm{\kappa}\cdot\bm{\xi}_{\perp}-\mu BJ_{0}\nabla\cdot\bm{\xi}_{\perp}\right]
\]
where $\bm{\kappa}$ stands for the curvature. The solution of the
drift kinetic equation can be written as

\[
\hat{g}=\frac{z}{M}\intop^{t}\frac{\partial f}{\partial\mathcal{E}}\left(\omega-\omega_{*}\right)\hat{X}dt'
\]
where the time integration is along the particle trajectory. Assuming
$S_{m}(t=0)=0$, one can write in terms of the following Fourier series
\[
\hat{X}=\sum_{m,l}e^{-i\omega t+i\overline{\omega}_{Dm}t}X_{ml}e^{il\omega_{\theta}t}
\]

\[
X_{ml}=\frac{1}{\tau_{b}}\oint dt'\hat{X}e^{i\intop^{t'}\left(\omega_{Dm}-\overline{\omega}_{Dm}-l\omega_{\theta}\right)dt''}
\]
$\Im\delta W_{k}$ can then be calculated from

\[
\delta W_{k}=-\int\int\hat{g}{\bf v}\cdot\left({\bf v}\cdot\nabla\right)\xi^{*}d{\bf r}_{c}d{\bf v}=-\frac{iz}{\omega}\int\int\hat{X}^{*}\hat{g}d{\bf r}_{c}d{\bf v}
\]
The phase-space integration is given by

\[
\begin{array}{c}
\intop d\Gamma...=\int d^{3}J\int d^{3}\Theta...=\left(2\pi\right)^{3}\int d^{3}J\underset{l}{\sum}...=\\
=\int d{\bf r}_{c}\int d{\bf v}...=\left(2\pi\right)^{2}\underset{\sigma_{\parallel}}{\sum}\frac{B}{\omega_{c}}\int dP_{\varphi}\int d\mathcal{E}\int d\mu\int dt...,
\end{array}
\]
with $dt$ being associated to the fast particle orbital motion. The
integrals over $\varphi$ and $\theta_{g}$ will contribute with $2\pi$
each. Note that in NOVA, $P_{\varphi}$, $\mathcal{E}$ and $\mu$
are defined without the mass. If they had usual units, we would have
$\int d{\bf r}_{c}\int d{\bf v}...=\left(2\pi\right)^{2}\left(\frac{c}{zM^{2}}\right)\underset{\sigma_{\parallel}}{\sum}\int dP_{\varphi}\int d\mathcal{E}\int d\mu\int dt...$
. We can write

\[
\hat{g}=\frac{iz}{M}\sum_{m,l}\frac{e^{-i\omega t+i\overline{\omega}_{Dm}t+il\omega_{\theta}t}}{\omega-\overline{\omega}_{Dm}-l\omega_{\theta}}\frac{\partial f}{\partial\mathcal{E}}\left(\omega-\omega_{*}\right)X_{ml}
\]
Therefore

\[
\delta W_{k}=-\frac{\left(2\pi M\right)^{2}c\omega}{z}\underset{\sigma_{\parallel}}{\sum}\int dP_{\varphi}d\mu d\mathcal{E}dt\sum_{m,m',l,l'}G_{m'l}^{*}\frac{\mathcal{E}^{2}\left(\frac{\partial f}{\partial\mathcal{E}}\right)\left(1-\nicefrac{\omega_{*}}{\omega}\right)}{\omega-\overline{\omega}_{Dm}-l\omega_{\theta}}G_{ml}e^{-it\left(\overline{\omega}_{Dm'}-\overline{\omega}_{Dm}+\left(l'-l\right)\omega_{\theta}\right)}
\]
where the ``drift'' frequency is $\omega_{Dm}\equiv\frac{dS_{m}}{dt}$
and $\overline{\omega}_{Dm}$ means orbit averaged drift frequency,
The matrix elements are defined as $G_{ml}=-i\frac{zX_{ml}}{\omega M\mathcal{E}}$.
The Plemelj formula could be used to develop the denominator. So,
the imaginary part of $\delta W_{k}$ becomes 

\[
\Im\delta W_{k}=\frac{\left(2M\right)^{2}\pi^{3}c\omega_{r}}{z}\int dP_{\varphi}d\mu d\mathcal{E}\sum_{m,m',l}G_{m'l}^{*}\mathcal{E}^{2}\tau_{b}\left(\frac{\partial f}{\partial\mathcal{E}}\right)\left(1-\nicefrac{\omega_{*}}{\omega_{r}}\right)G_{ml}\delta\left(\omega_{r}-\overline{\omega}_{D0}-l\omega_{\theta}\right)
\]
The growth rate in NOVA is then given by:
\[
\gamma=\frac{\left(2M\right)^{2}\pi^{3}c\omega^{2}}{2z\omega^{2}\int\rho\left|\bm{\xi}\right|^{2}d{\bf r}}\underset{\sigma_{\parallel}}{\sum}\int dP_{\varphi}d\mu d\mathcal{E}\sum_{m,m',l}G_{m'l}^{*}\mathcal{E}^{2}\tau_{b}\left(\frac{\partial f}{\partial\mathcal{E}}-\frac{n}{\omega}\frac{\partial f}{\partial P_{\varphi}}\right)G_{ml}\delta\left(\omega-\overline{\omega}_{D0}-l\omega_{\theta}\right)
\]
Now we want to compare the growth rate in NOVA and in Kaufman's normal
mode theory, in order to be able to relate their respective mode structure
information ($G$ and $\alpha_{{\bf l}}$) and build a quasilinear
theory based on that. Using $\delta K=\omega^{2}\int\rho\left|\bm{\xi}\right|^{2}d{\bf r}$,
the comparison leads to $\frac{M\mathcal{E}\omega}{\sqrt{2}}\sum_{m}G_{ml}\equiv\int d^{3}x\mathbf{e}\left(\mathbf{x},\omega\right)\cdot{\bf j_{l}}\left({\bf x}\mid{\bf J}\right)=\frac{\omega}{i}V_{{\bf l}}$
or, alternatively
\begin{equation}
\alpha_{{\bf l}}\left({\bf J}\right)=\left(\frac{M\mathcal{E}\omega}{\sqrt{2}}\right)^{2}\sum_{m,m'}G_{m'l}^{*}G_{ml}=\omega^{2}\left|V_{{\bf l}}\right|^{2}\label{eq:square}
\end{equation}
where
\[
V_{l}=\frac{iz}{\omega}\int\frac{d^{3}\Theta}{(2\pi)^{3}}\mathbf{e}\left(\mathbf{r},\omega\right)\cdot{\bf v}\left({\bf J},\bm{\Theta}\right)e^{-i\mathbf{l}\cdot\bm{\Theta}}
\]
which is consistent with the expression given in \cite{BerkPPR97}.
The mode structure is provided in NOVA via the $G$ matrices. They
are calculated via time integration that account for the eigenmode
felt by a particle in a given trajectory. A weighted integration is
performing according to how much time a particle spends at each position
of its trajectory. Mirror-trapped particles have their parallel velocity
reversed at the tips of a banana orbit, where they tend to remain
longer and consequently the local mode structure will have an important
contribution for the overall integration. By using equation \eqref{eq:square}
we can now express the quasilinear diffusion equation, \eqref{eq:QLeq}
and \eqref{eq:QLeqDiffusionCoeff}, in terms of NOVA code notation.

\section{Expected saturation levels from a single mode perturbation theory}

\label{sec:expsatlevl}For a single, isolated mode in a simplified
bump-on-tail situation, at the saturation ($\gamma_{L}\simeq\gamma_{d}$,
$\partial f/\partial t=0$), we have
\[
\frac{\pi}{2}\omega_{b}^{4}\mathcal{F}\frac{\partial f}{\partial\Omega}+\nu_{scatt}^{3}\frac{\partial(f-f_{0})}{\partial\Omega}=0.
\]
Analytic results far from and close to marginal stability can be used
to lead the choice of the broadening parameters \cite{GhantousPoP2014,GhantousThesis}.
For the case of a flat-topped window function, $\mathcal{F}=1/\Delta\Omega$,
we have
\begin{equation}
\Delta\Omega=\frac{\pi}{2}\frac{\omega_{b}^{4}}{\nu_{scatt}^{3}}\frac{\gamma_{d}}{\gamma_{L,0}-\gamma_{d}}.\label{eq:DeltaOmegaSaturation}
\end{equation}

\subsubsection*{Near marginal stability}

For this case $\Delta\Omega\approx c\nu_{scatt}$. The expected saturation
level from analytical theory for the case $\omega_{b}/\nu_{scatt}\ll1$
to be \cite{BerkPRL96,GorelenkovPoP99nlin} 

\[
\omega_{b}\simeq1.18\nu_{scatt}\left(\frac{\gamma_{L}}{\gamma_{d}}-1\right)^{1/4},
\]
which, when substituted in \eqref{eq:DeltaOmegaSaturation} leads
to $a=2.7$.

\subsubsection*{Far from marginal stability}

For this case $\Delta\Omega\approx a\omega_{b}$. The expected saturation
level from analytical theory for the case $\omega_{b}/\nu_{scatt}\gg1$
is \cite{BerkBreizman1990c,GorelenkovPoP99nlin}

\[
\omega_{b}\simeq1.2\nu_{scatt}\left(\frac{\gamma_{L}}{\gamma_{d}}\right)^{1/3},
\]
which, when substituted in \eqref{eq:DeltaOmegaSaturation} leads
to $c=2.5$. There are no known analytical solutions to rely on in
order to determine the constant $b$. It was simply taken equal to
$c$ in previous works \cite{GhantousPoP2014,GhantousThesis,BerkPoP96,BerkNF95}.
We have verified that the saturation levels obtained numerically with
RBQ are not sensitive with respect to the exact value of $b$. This
is because as the mode starts its growth, the term $c\nu_{scatt}$
has been observed to be the dominant broadening component for practical
cases, as inferred by NOVA-K. On he other hand, as the mode approaches
saturation, the term $a\omega_{b,n}$ outweights $b\left|\gamma_{L,n}-\gamma_{d,n}\right|$
since the latter approaches zero. The study of the parametric dependencies
of the broadened window for realistic eigenmodes is under way \cite{MengGuoIAEA2017}
using the guiding-center, particle following code ORBIT \cite{WhitePoP84}. 

\bibliographystyle{apsrev4-1}

\begin{thebibliography}{41}%
\makeatletter
\providecommand \@ifxundefined [1]{%
 \@ifx{#1\undefined}
}%
\providecommand \@ifnum [1]{%
 \ifnum #1\expandafter \@firstoftwo
 \else \expandafter \@secondoftwo
 \fi
}%
\providecommand \@ifx [1]{%
 \ifx #1\expandafter \@firstoftwo
 \else \expandafter \@secondoftwo
 \fi
}%
\providecommand \natexlab [1]{#1}%
\providecommand \enquote  [1]{``#1''}%
\providecommand \bibnamefont  [1]{#1}%
\providecommand \bibfnamefont [1]{#1}%
\providecommand \citenamefont [1]{#1}%
\providecommand \href@noop [0]{\@secondoftwo}%
\providecommand \href [0]{\begingroup \@sanitize@url \@href}%
\providecommand \@href[1]{\@@startlink{#1}\@@href}%
\providecommand \@@href[1]{\endgroup#1\@@endlink}%
\providecommand \@sanitize@url [0]{\catcode `\\12\catcode `\$12\catcode
  `\&12\catcode `\#12\catcode `\^12\catcode `\_12\catcode `\%12\relax}%
\providecommand \@@startlink[1]{}%
\providecommand \@@endlink[0]{}%
\providecommand \url  [0]{\begingroup\@sanitize@url \@url }%
\providecommand \@url [1]{\endgroup\@href {#1}{\urlprefix }}%
\providecommand \urlprefix  [0]{URL }%
\providecommand \Eprint [0]{\href }%
\providecommand \doibase [0]{http://dx.doi.org/}%
\providecommand \selectlanguage [0]{\@gobble}%
\providecommand \bibinfo  [0]{\@secondoftwo}%
\providecommand \bibfield  [0]{\@secondoftwo}%
\providecommand \translation [1]{[#1]}%
\providecommand \BibitemOpen [0]{}%
\providecommand \bibitemStop [0]{}%
\providecommand \bibitemNoStop [0]{.\EOS\space}%
\providecommand \EOS [0]{\spacefactor3000\relax}%
\providecommand \BibitemShut  [1]{\csname bibitem#1\endcsname}%
\let\auto@bib@innerbib\@empty
\bibitem [{\citenamefont {Gorelenkov}\ \emph {et~al.}(2014)\citenamefont
  {Gorelenkov}, \citenamefont {Pinches},\ and\ \citenamefont
  {Toi}}]{GorelenkovNF14rev}%
  \BibitemOpen
  \bibfield  {author} {\bibinfo {author} {\bibfnamefont {N.~N.}\ \bibnamefont
  {Gorelenkov}}, \bibinfo {author} {\bibfnamefont {S.~D.}\ \bibnamefont
  {Pinches}}, \ and\ \bibinfo {author} {\bibfnamefont {K.}~\bibnamefont
  {Toi}},\ }\href@noop {} {\bibfield  {journal} {\bibinfo  {journal} {Nucl.
  Fusion}\ }\textbf {\bibinfo {volume} {54}},\ \bibinfo {pages} {125001}
  (\bibinfo {year} {2014})}\BibitemShut {NoStop}%
\bibitem [{\citenamefont {Duarte}(2017)}]{DuartePhD17}%
  \BibitemOpen
  \bibfield  {author} {\bibinfo {author} {\bibfnamefont {V.~N.}\ \bibnamefont
  {Duarte}},\ }\emph {\bibinfo {title} {Quasilinear and nonlinear dynamics of
  energetic-ion-driven \protect{Alfv\'{e}}n eigenmodes,
  http://www.teses.usp.br/teses/disponiveis/43/43134/tde-01082017-195849/}},\
  \href@noop {} {Ph.D. thesis},\ \bibinfo  {school} {University of S\~{a}o
  Paulo}, \bibinfo {address} {S\~{a}o Paulo, Brazil} (\bibinfo {year}
  {2017})\BibitemShut {NoStop}%
\bibitem [{\citenamefont {Collins}\ \emph {et~al.}(2016)\citenamefont
  {Collins}, \citenamefont {Heidbrink}, \citenamefont {Austin}, \citenamefont
  {Kramer}, \citenamefont {Pace}, \citenamefont {Petty}, \citenamefont
  {Stagner}, \citenamefont {\protect{Van} Zeeland}, \citenamefont {White},
  \citenamefont {Zhu},\ and\ \citenamefont {\protect{The DIII-D
  team}}}]{CollinsPRL16}%
  \BibitemOpen
  \bibfield  {author} {\bibinfo {author} {\bibfnamefont {C.~S.}\ \bibnamefont
  {Collins}}, \bibinfo {author} {\bibfnamefont {W.~W.}\ \bibnamefont
  {Heidbrink}}, \bibinfo {author} {\bibfnamefont {M.~E.}\ \bibnamefont
  {Austin}}, \bibinfo {author} {\bibfnamefont {G.~J.}\ \bibnamefont {Kramer}},
  \bibinfo {author} {\bibfnamefont {D.~C.}\ \bibnamefont {Pace}}, \bibinfo
  {author} {\bibfnamefont {C.~C.}\ \bibnamefont {Petty}}, \bibinfo {author}
  {\bibfnamefont {L.}~\bibnamefont {Stagner}}, \bibinfo {author} {\bibfnamefont
  {M.~A.}\ \bibnamefont {\protect{Van} Zeeland}}, \bibinfo {author}
  {\bibfnamefont {R.~B.}\ \bibnamefont {White}}, \bibinfo {author}
  {\bibfnamefont {Y.~B.}\ \bibnamefont {Zhu}}, \ and\ \bibinfo {author}
  {\bibnamefont {\protect{The DIII-D team}}},\ }\href {\doibase
  10.1103/PhysRevLett.116.095001} {\bibfield  {journal} {\bibinfo  {journal}
  {Phys. Rev. Letters}\ }\textbf {\bibinfo {volume} {116}},\ \bibinfo {pages}
  {095001} (\bibinfo {year} {2016})}\BibitemShut {NoStop}%
\bibitem [{\citenamefont {Heidbrink}\ \emph {et~al.}(2008)\citenamefont
  {Heidbrink}, \citenamefont {\protect{Van~Zeeland}}, \citenamefont {Austin},
  \citenamefont {Burrell}, \citenamefont {Gorelenkov}, \citenamefont {Kramer},
  \citenamefont {Luo}, \citenamefont {Makowski}, \citenamefont {McKee},
  \citenamefont {Muscatello}, \citenamefont {Nazikian}, \citenamefont {Ruskov},
  \citenamefont {Solomon}, \citenamefont {White},\ and\ \citenamefont
  {Zhu}}]{HeidbrinkNF08}%
  \BibitemOpen
  \bibfield  {author} {\bibinfo {author} {\bibfnamefont {W.~W.}\ \bibnamefont
  {Heidbrink}}, \bibinfo {author} {\bibfnamefont {M.~A.}\ \bibnamefont
  {\protect{Van~Zeeland}}}, \bibinfo {author} {\bibfnamefont {M.~E.}\
  \bibnamefont {Austin}}, \bibinfo {author} {\bibfnamefont {K.~H.}\
  \bibnamefont {Burrell}}, \bibinfo {author} {\bibfnamefont {N.~N.}\
  \bibnamefont {Gorelenkov}}, \bibinfo {author} {\bibfnamefont {G.~J.}\
  \bibnamefont {Kramer}}, \bibinfo {author} {\bibfnamefont {Y.}~\bibnamefont
  {Luo}}, \bibinfo {author} {\bibfnamefont {M.~A.}\ \bibnamefont {Makowski}},
  \bibinfo {author} {\bibfnamefont {G.~R.}\ \bibnamefont {McKee}}, \bibinfo
  {author} {\bibfnamefont {C.}~\bibnamefont {Muscatello}}, \bibinfo {author}
  {\bibfnamefont {R.}~\bibnamefont {Nazikian}}, \bibinfo {author}
  {\bibfnamefont {E.}~\bibnamefont {Ruskov}}, \bibinfo {author} {\bibfnamefont
  {W.~M.}\ \bibnamefont {Solomon}}, \bibinfo {author} {\bibfnamefont {R.~B.}\
  \bibnamefont {White}}, \ and\ \bibinfo {author} {\bibfnamefont
  {Y.}~\bibnamefont {Zhu}},\ }\href@noop {} {\bibfield  {journal} {\bibinfo
  {journal} {Nucl. Fusion}\ }\textbf {\bibinfo {volume} {48}},\ \bibinfo
  {pages} {084001} (\bibinfo {year} {2008})}\BibitemShut {NoStop}%
\bibitem [{\citenamefont {Drummond}\ and\ \citenamefont
  {Pines}(1962)}]{Drummond_Pines_1962}%
  \BibitemOpen
  \bibfield  {author} {\bibinfo {author} {\bibfnamefont {W.}~\bibnamefont
  {Drummond}}\ and\ \bibinfo {author} {\bibfnamefont {D.}~\bibnamefont
  {Pines}},\ }\href@noop {} {\bibfield  {journal} {\bibinfo  {journal} {Nucl.
  Fusion}\ }\textbf {\bibinfo {volume} {Suppl. 2, Pt. 3}} (\bibinfo {year}
  {1962})}\BibitemShut {NoStop}%
\bibitem [{\citenamefont {Vedenov}\ \emph {et~al.}(1961)\citenamefont
  {Vedenov}, \citenamefont {Velikhov},\ and\ \citenamefont
  {Sagdeev}}]{VedenovSagdeev1961}%
  \BibitemOpen
  \bibfield  {author} {\bibinfo {author} {\bibfnamefont {A.~A.}\ \bibnamefont
  {Vedenov}}, \bibinfo {author} {\bibfnamefont {E.~P.}\ \bibnamefont
  {Velikhov}}, \ and\ \bibinfo {author} {\bibfnamefont {R.~Z.}\ \bibnamefont
  {Sagdeev}},\ }\href {http://stacks.iop.org/0038-5670/4/i=2/a=A12} {\bibfield
  {journal} {\bibinfo  {journal} {Sov. Phys. Uspekhi}\ }\textbf {\bibinfo
  {volume} {4}},\ \bibinfo {pages} {332} (\bibinfo {year} {1961})}\BibitemShut
  {NoStop}%
\bibitem [{\citenamefont {Duarte}\ \emph
  {et~al.}(2017{\natexlab{a}})\citenamefont {Duarte}, \citenamefont {Berk},
  \citenamefont {Gorelenkov}, \citenamefont {Heidbrink}, \citenamefont
  {Kramer}, \citenamefont {Nazikian}, \citenamefont {Pace}, \citenamefont
  {Podest\protect{\`{a}}}, \citenamefont {Tobias},\ and\ \citenamefont
  {\protect{Van Zeeland}}}]{DuarteAxivPRL}%
  \BibitemOpen
  \bibfield  {author} {\bibinfo {author} {\bibfnamefont {V.~N.}\ \bibnamefont
  {Duarte}}, \bibinfo {author} {\bibfnamefont {H.~L.}\ \bibnamefont {Berk}},
  \bibinfo {author} {\bibfnamefont {N.~N.}\ \bibnamefont {Gorelenkov}},
  \bibinfo {author} {\bibfnamefont {W.~W.}\ \bibnamefont {Heidbrink}}, \bibinfo
  {author} {\bibfnamefont {G.~J.}\ \bibnamefont {Kramer}}, \bibinfo {author}
  {\bibfnamefont {R.}~\bibnamefont {Nazikian}}, \bibinfo {author}
  {\bibfnamefont {D.~C.}\ \bibnamefont {Pace}}, \bibinfo {author}
  {\bibfnamefont {M.}~\bibnamefont {Podest\protect{\`{a}}}}, \bibinfo {author}
  {\bibfnamefont {B.~J.}\ \bibnamefont {Tobias}}, \ and\ \bibinfo {author}
  {\bibfnamefont {M.~A.}\ \bibnamefont {\protect{Van Zeeland}}},\ }\href
  {http://stacks.iop.org/0029-5515/57/i=5/a=054001} {\bibfield  {journal}
  {\bibinfo  {journal} {Nuclear Fusion}\ }\textbf {\bibinfo {volume} {57}},\
  \bibinfo {pages} {054001} (\bibinfo {year} {2017}{\natexlab{a}})}\BibitemShut
  {NoStop}%
\bibitem [{\citenamefont {Duarte}\ \emph
  {et~al.}(2017{\natexlab{b}})\citenamefont {Duarte}, \citenamefont {Berk},
  \citenamefont {Gorelenkov}, \citenamefont {Heidbrink}, \citenamefont
  {Kramer}, \citenamefont {Nazikian}, \citenamefont {Pace}, \citenamefont
  {Podestà},\ and\ \citenamefont {Zeeland}}]{DuartePoP2017}%
  \BibitemOpen
  \bibfield  {author} {\bibinfo {author} {\bibfnamefont {V.~N.}\ \bibnamefont
  {Duarte}}, \bibinfo {author} {\bibfnamefont {H.~L.}\ \bibnamefont {Berk}},
  \bibinfo {author} {\bibfnamefont {N.~N.}\ \bibnamefont {Gorelenkov}},
  \bibinfo {author} {\bibfnamefont {W.~W.}\ \bibnamefont {Heidbrink}}, \bibinfo
  {author} {\bibfnamefont {G.~J.}\ \bibnamefont {Kramer}}, \bibinfo {author}
  {\bibfnamefont {R.}~\bibnamefont {Nazikian}}, \bibinfo {author}
  {\bibfnamefont {D.~C.}\ \bibnamefont {Pace}}, \bibinfo {author}
  {\bibfnamefont {M.}~\bibnamefont {Podestà}}, \ and\ \bibinfo {author}
  {\bibfnamefont {M.~A.~V.}\ \bibnamefont {Zeeland}},\ }\href {\doibase
  10.1063/1.5007811} {\bibfield  {journal} {\bibinfo  {journal} {Physics of
  Plasmas}\ }\textbf {\bibinfo {volume} {24}},\ \bibinfo {pages} {122508}
  (\bibinfo {year} {2017}{\natexlab{b}})},\ \Eprint
  {http://arxiv.org/abs/https://doi.org/10.1063/1.5007811}
  {https://doi.org/10.1063/1.5007811} \BibitemShut {NoStop}%
\bibitem [{\citenamefont {Kaufman}(1972)}]{KaufmanQLPoF1972}%
  \BibitemOpen
  \bibfield  {author} {\bibinfo {author} {\bibfnamefont {A.~N.}\ \bibnamefont
  {Kaufman}},\ }\href {\doibase http://dx.doi.org/10.1063/1.1694031} {\bibfield
   {journal} {\bibinfo  {journal} {Phys. Fluids}\ }\textbf {\bibinfo {volume}
  {15}},\ \bibinfo {eid} {1063} (\bibinfo {year} {1972})}\BibitemShut {NoStop}%
\bibitem [{\citenamefont {Chirikov}(1960)}]{chirikov1960resonance}%
  \BibitemOpen
  \bibfield  {author} {\bibinfo {author} {\bibfnamefont {B.}~\bibnamefont
  {Chirikov}},\ }\href@noop {} {\bibfield  {journal} {\bibinfo  {journal}
  {Journal of Nuclear Energy. Part C, Plasma Physics, Accelerators,
  Thermonuclear Research}\ }\textbf {\bibinfo {volume} {1}},\ \bibinfo {pages}
  {253} (\bibinfo {year} {1960})}\BibitemShut {NoStop}%
\bibitem [{\citenamefont {Berk}\ \emph {et~al.}(1995)\citenamefont {Berk},
  \citenamefont {Breizman}, \citenamefont {Fitzpatrick},\ and\ \citenamefont
  {Wong}}]{BerkNF95}%
  \BibitemOpen
  \bibfield  {author} {\bibinfo {author} {\bibfnamefont {H.~L.}\ \bibnamefont
  {Berk}}, \bibinfo {author} {\bibfnamefont {B.~N.}\ \bibnamefont {Breizman}},
  \bibinfo {author} {\bibfnamefont {J.}~\bibnamefont {Fitzpatrick}}, \ and\
  \bibinfo {author} {\bibfnamefont {H.~V.}\ \bibnamefont {Wong}},\ }\href
  {http://stacks.iop.org/0029-5515/35/i=12/a=I30} {\bibfield  {journal}
  {\bibinfo  {journal} {Nucl. Fusion}\ }\textbf {\bibinfo {volume} {35}},\
  \bibinfo {pages} {1661} (\bibinfo {year} {1995})}\BibitemShut {NoStop}%
\bibitem [{\citenamefont {Berk}\ \emph
  {et~al.}(1996{\natexlab{a}})\citenamefont {Berk}, \citenamefont {Breizman},
  \citenamefont {Fitzpatrick}, \citenamefont {Pekker}, \citenamefont {Wong},\
  and\ \citenamefont {Wong}}]{BerkPoP96}%
  \BibitemOpen
  \bibfield  {author} {\bibinfo {author} {\bibfnamefont {H.~L.}\ \bibnamefont
  {Berk}}, \bibinfo {author} {\bibfnamefont {B.~N.}\ \bibnamefont {Breizman}},
  \bibinfo {author} {\bibfnamefont {J.}~\bibnamefont {Fitzpatrick}}, \bibinfo
  {author} {\bibfnamefont {M.~S.}\ \bibnamefont {Pekker}}, \bibinfo {author}
  {\bibfnamefont {H.~V.}\ \bibnamefont {Wong}}, \ and\ \bibinfo {author}
  {\bibfnamefont {K.~L.}\ \bibnamefont {Wong}},\ }\href {\doibase
  http://dx.doi.org/10.1063/1.871978} {\bibfield  {journal} {\bibinfo
  {journal} {Phys. Plasmas}\ }\textbf {\bibinfo {volume} {3}},\ \bibinfo
  {pages} {1827} (\bibinfo {year} {1996}{\natexlab{a}})}\BibitemShut {NoStop}%
\bibitem [{\citenamefont {Meng}\ \emph {et~al.}(2018)\citenamefont {Meng},
  \citenamefont {Gorelenkov}, \citenamefont {Duarte}, \citenamefont {Berk},
  \citenamefont {White},\ and\ \citenamefont {Wang}}]{MengGuoIAEA2017}%
  \BibitemOpen
  \bibfield  {author} {\bibinfo {author} {\bibfnamefont {G.}~\bibnamefont
  {Meng}}, \bibinfo {author} {\bibfnamefont {N.~N.}\ \bibnamefont
  {Gorelenkov}}, \bibinfo {author} {\bibfnamefont {V.~N.}\ \bibnamefont
  {Duarte}}, \bibinfo {author} {\bibfnamefont {H.~L.}\ \bibnamefont {Berk}},
  \bibinfo {author} {\bibfnamefont {R.~B.}\ \bibnamefont {White}}, \ and\
  \bibinfo {author} {\bibfnamefont {X.}~\bibnamefont {Wang}},\ }\href {\doibase
  10.1088/1741-4326/aaa918} {\bibfield  {journal} {\bibinfo  {journal} {Nuclear
  Fusion}\ } (\bibinfo {year} {2018}),\ 10.1088/1741-4326/aaa918}\BibitemShut
  {NoStop}%
\bibitem [{\citenamefont {Fitzpatrick}(1997)}]{FitzpatrickPhD97}%
  \BibitemOpen
  \bibfield  {author} {\bibinfo {author} {\bibfnamefont {J.}~\bibnamefont
  {Fitzpatrick}},\ }\emph {\bibinfo {title} {A Numerical Model of Wave-Induced
  Fast Particle Transport in a Fusion Plasma}},\ \href@noop {} {Ph.D. thesis},\
  \bibinfo  {school} {University of California, Berkeley}, \bibinfo {address}
  {Berkeley, CA} (\bibinfo {year} {1997})\BibitemShut {NoStop}%
\bibitem [{\citenamefont {Ghantous}\ \emph
  {et~al.}(2014{\natexlab{a}})\citenamefont {Ghantous}, \citenamefont {Berk},\
  and\ \citenamefont {Gorelenkov}}]{GhantousPoP2014}%
  \BibitemOpen
  \bibfield  {author} {\bibinfo {author} {\bibfnamefont {K.}~\bibnamefont
  {Ghantous}}, \bibinfo {author} {\bibfnamefont {H.~L.}\ \bibnamefont {Berk}},
  \ and\ \bibinfo {author} {\bibfnamefont {N.~N.}\ \bibnamefont {Gorelenkov}},\
  }\href {\doibase http://dx.doi.org/10.1063/1.4869242} {\bibfield  {journal}
  {\bibinfo  {journal} {Phys. Plasmas}\ }\textbf {\bibinfo {volume} {21}},\
  \bibinfo {eid} {032119} (\bibinfo {year} {2014}{\natexlab{a}}),\
  http://dx.doi.org/10.1063/1.4869242}\BibitemShut {NoStop}%
\bibitem [{\citenamefont {Ghantous}(2013)}]{GhantousThesis}%
  \BibitemOpen
  \bibfield  {author} {\bibinfo {author} {\bibfnamefont {K.}~\bibnamefont
  {Ghantous}},\ }\emph {\bibinfo {title} {Reduced Quasilinear Models for
  Energetic Particles Interaction with \protect{Alfv\'{e}}n Eigenmodes}},\
  \href@noop {} {Ph.D. thesis},\ \bibinfo  {school} {Princeton University}
  (\bibinfo {year} {2013})\BibitemShut {NoStop}%
\bibitem [{\citenamefont {Heidbrink}\ \emph {et~al.}(2017)\citenamefont
  {Heidbrink}, \citenamefont {Collins}, \citenamefont {Podest\protect{\`{a}}},
  \citenamefont {Kramer}, \citenamefont {Pace}, \citenamefont {Petty},
  \citenamefont {Stagner}, \citenamefont {\protect{Van Zeeland}}, \citenamefont
  {White},\ and\ \citenamefont {Zhu}}]{HeidbrinkPOP17aecgm}%
  \BibitemOpen
  \bibfield  {author} {\bibinfo {author} {\bibfnamefont {W.~W.}\ \bibnamefont
  {Heidbrink}}, \bibinfo {author} {\bibfnamefont {C.~S.}\ \bibnamefont
  {Collins}}, \bibinfo {author} {\bibfnamefont {M.}~\bibnamefont
  {Podest\protect{\`{a}}}}, \bibinfo {author} {\bibfnamefont {G.~J.}\
  \bibnamefont {Kramer}}, \bibinfo {author} {\bibfnamefont {D.~C.}\
  \bibnamefont {Pace}}, \bibinfo {author} {\bibfnamefont {C.~C.}\ \bibnamefont
  {Petty}}, \bibinfo {author} {\bibfnamefont {L.}~\bibnamefont {Stagner}},
  \bibinfo {author} {\bibfnamefont {M.~A.}\ \bibnamefont {\protect{Van
  Zeeland}}}, \bibinfo {author} {\bibfnamefont {R.~B.}\ \bibnamefont {White}},
  \ and\ \bibinfo {author} {\bibfnamefont {Y.~B.}\ \bibnamefont {Zhu}},\
  }\href@noop {} {\bibfield  {journal} {\bibinfo  {journal} {Phys. Plasmas}\
  }\textbf {\bibinfo {volume} {24}},\ \bibinfo {pages} {056109} (\bibinfo
  {year} {2017})}\BibitemShut {NoStop}%
\bibitem [{\citenamefont {Podest\protect{\`{a}}}\ \emph
  {et~al.}(2014)\citenamefont {Podest\protect{\`{a}}}, \citenamefont
  {Gorelenkova},\ and\ \citenamefont {White}}]{PodestaPPCF14}%
  \BibitemOpen
  \bibfield  {author} {\bibinfo {author} {\bibfnamefont {M.}~\bibnamefont
  {Podest\protect{\`{a}}}}, \bibinfo {author} {\bibfnamefont {M.~V.}\
  \bibnamefont {Gorelenkova}}, \ and\ \bibinfo {author} {\bibfnamefont {R.~B.}\
  \bibnamefont {White}},\ }\href@noop {} {\bibfield  {journal} {\bibinfo
  {journal} {Plasma Phys. Control. Fusion}\ }\textbf {\bibinfo {volume} {56}},\
  \bibinfo {pages} {055003} (\bibinfo {year} {2014})}\BibitemShut {NoStop}%
\bibitem [{\citenamefont {Podest\`{a}}\ \emph {et~al.}(2017)\citenamefont
  {Podest\`{a}}, \citenamefont {Gorelenkova}, \citenamefont {Gorelenkov},\ and\
  \citenamefont {White}}]{PodestaPPCF17}%
  \BibitemOpen
  \bibfield  {author} {\bibinfo {author} {\bibfnamefont {M.}~\bibnamefont
  {Podest\`{a}}}, \bibinfo {author} {\bibfnamefont {M.}~\bibnamefont
  {Gorelenkova}}, \bibinfo {author} {\bibfnamefont {N.~N.}\ \bibnamefont
  {Gorelenkov}}, \ and\ \bibinfo {author} {\bibfnamefont {R.~B.}\ \bibnamefont
  {White}},\ }\href@noop {} {\bibfield  {journal} {\bibinfo  {journal} {Plasma
  Phys. Control. Fusion}\ }\textbf {\bibinfo {volume} {59}},\ \bibinfo {pages}
  {095008} (\bibinfo {year} {2017})}\BibitemShut {NoStop}%
\bibitem [{\citenamefont {Goldston}\ \emph {et~al.}(1981)\citenamefont
  {Goldston}, \citenamefont {Mc\protect{C}une}, \citenamefont {Towner},
  \citenamefont {Davis}, \citenamefont {Hawryluk},\ and\ \citenamefont
  {Schmidt}}]{GoldstonJCP81}%
  \BibitemOpen
  \bibfield  {author} {\bibinfo {author} {\bibfnamefont {R.~J.}\ \bibnamefont
  {Goldston}}, \bibinfo {author} {\bibfnamefont {D.~C.}\ \bibnamefont
  {Mc\protect{C}une}}, \bibinfo {author} {\bibfnamefont {H.~H.}\ \bibnamefont
  {Towner}}, \bibinfo {author} {\bibfnamefont {S.~L.}\ \bibnamefont {Davis}},
  \bibinfo {author} {\bibfnamefont {R.~J.}\ \bibnamefont {Hawryluk}}, \ and\
  \bibinfo {author} {\bibfnamefont {G.~L.}\ \bibnamefont {Schmidt}},\ }\href
  {\doibase http://dx.doi.org/10.1016/0021-9991(81)90111-X} {\bibfield
  {journal} {\bibinfo  {journal} {J. Comput. Phys.}\ }\textbf {\bibinfo
  {volume} {43}},\ \bibinfo {pages} {61} (\bibinfo {year} {1981})}\BibitemShut
  {NoStop}%
\bibitem [{\citenamefont {Dupree}(1966)}]{Dupree1966}%
  \BibitemOpen
  \bibfield  {author} {\bibinfo {author} {\bibfnamefont {T.~H.}\ \bibnamefont
  {Dupree}},\ }\href {\doibase 10.1063/1.1761932} {\bibfield  {journal}
  {\bibinfo  {journal} {Physics of Fluids}\ }\textbf {\bibinfo {volume} {9}},\
  \bibinfo {pages} {1773} (\bibinfo {year} {1966})}\BibitemShut {NoStop}%
\bibitem [{\citenamefont {White}\ \emph {et~al.}(2010)\citenamefont {White},
  \citenamefont {Gorelenkov}, \citenamefont {Heidbrink},\ and\ \citenamefont
  {\protect{Van~Zeeland}}}]{WhitePPCF10}%
  \BibitemOpen
  \bibfield  {author} {\bibinfo {author} {\bibfnamefont {R.~B.}\ \bibnamefont
  {White}}, \bibinfo {author} {\bibfnamefont {N.~N.}\ \bibnamefont
  {Gorelenkov}}, \bibinfo {author} {\bibfnamefont {W.~W.}\ \bibnamefont
  {Heidbrink}}, \ and\ \bibinfo {author} {\bibfnamefont {M.~A.}\ \bibnamefont
  {\protect{Van~Zeeland}}},\ }\href@noop {} {\bibfield  {journal} {\bibinfo
  {journal} {Plasma Phys. Control. Fusion}\ }\textbf {\bibinfo {volume} {53}},\
  \bibinfo {pages} {045012} (\bibinfo {year} {2010})}\BibitemShut {NoStop}%
\bibitem [{\citenamefont {Berk}\ and\ \citenamefont
  {Breizman}(1990{\natexlab{a}})}]{BerkBreizman1990b}%
  \BibitemOpen
  \bibfield  {author} {\bibinfo {author} {\bibfnamefont {H.~L.}\ \bibnamefont
  {Berk}}\ and\ \bibinfo {author} {\bibfnamefont {B.~N.}\ \bibnamefont
  {Breizman}},\ }\href {\doibase 10.1063/1.859405} {\bibfield  {journal}
  {\bibinfo  {journal} {Physics of Fluids B: Plasma Physics}\ }\textbf
  {\bibinfo {volume} {2}},\ \bibinfo {pages} {2235} (\bibinfo {year}
  {1990}{\natexlab{a}})},\ \Eprint
  {http://arxiv.org/abs/http://dx.doi.org/10.1063/1.859405}
  {http://dx.doi.org/10.1063/1.859405} \BibitemShut {NoStop}%
\bibitem [{\citenamefont {Berk}\ \emph {et~al.}(1997)\citenamefont {Berk},
  \citenamefont {Breizman},\ and\ \citenamefont {Pekker}}]{BerkPPR97}%
  \BibitemOpen
  \bibfield  {author} {\bibinfo {author} {\bibfnamefont {H.~L.}\ \bibnamefont
  {Berk}}, \bibinfo {author} {\bibfnamefont {B.~N.}\ \bibnamefont {Breizman}},
  \ and\ \bibinfo {author} {\bibfnamefont {M.~S.}\ \bibnamefont {Pekker}},\
  }\href@noop {} {\bibfield  {journal} {\bibinfo  {journal} {Plasma Phys.
  Reports}\ }\textbf {\bibinfo {volume} {23}},\ \bibinfo {pages} {778}
  (\bibinfo {year} {1997})}\BibitemShut {NoStop}%
\bibitem [{\citenamefont {Gorelenkov}\ \emph
  {et~al.}(1999{\natexlab{a}})\citenamefont {Gorelenkov}, \citenamefont {Chen},
  \citenamefont {White},\ and\ \citenamefont {Berk}}]{GorelenkovPoP99nlin}%
  \BibitemOpen
  \bibfield  {author} {\bibinfo {author} {\bibfnamefont {N.~N.}\ \bibnamefont
  {Gorelenkov}}, \bibinfo {author} {\bibfnamefont {Y.}~\bibnamefont {Chen}},
  \bibinfo {author} {\bibfnamefont {R.~B.}\ \bibnamefont {White}}, \ and\
  \bibinfo {author} {\bibfnamefont {H.~L.}\ \bibnamefont {Berk}},\ }\href@noop
  {} {\bibfield  {journal} {\bibinfo  {journal} {Phys. Plasmas}\ }\textbf
  {\bibinfo {volume} {6}},\ \bibinfo {pages} {629} (\bibinfo {year}
  {1999}{\natexlab{a}})}\BibitemShut {NoStop}%
\bibitem [{\citenamefont {Ghantous}\ \emph
  {et~al.}(2014{\natexlab{b}})\citenamefont {Ghantous}, \citenamefont {Berk},\
  and\ \citenamefont {Gorelenkov}}]{GhantousPoP14}%
  \BibitemOpen
  \bibfield  {author} {\bibinfo {author} {\bibfnamefont {K.}~\bibnamefont
  {Ghantous}}, \bibinfo {author} {\bibfnamefont {H.~L.}\ \bibnamefont {Berk}},
  \ and\ \bibinfo {author} {\bibfnamefont {N.~N.}\ \bibnamefont {Gorelenkov}},\
  }\href@noop {} {\bibfield  {journal} {\bibinfo  {journal} {Phys. Plasmas}\
  }\textbf {\bibinfo {volume} {21}},\ \bibinfo {pages} {032119} (\bibinfo
  {year} {2014}{\natexlab{b}})}\BibitemShut {NoStop}%
\bibitem [{\citenamefont {Pankin}\ \emph {et~al.}(2004)\citenamefont {Pankin},
  \citenamefont {McCune}, \citenamefont {Andre},\ and\ \citenamefont
  {et.al.}}]{PankinCPC04}%
  \BibitemOpen
  \bibfield  {author} {\bibinfo {author} {\bibfnamefont {A.}~\bibnamefont
  {Pankin}}, \bibinfo {author} {\bibfnamefont {D.}~\bibnamefont {McCune}},
  \bibinfo {author} {\bibfnamefont {R.}~\bibnamefont {Andre}}, \ and\ \bibinfo
  {author} {\bibnamefont {et.al.}},\ }\href@noop {} {\bibfield  {journal}
  {\bibinfo  {journal} {Comp. Phys. Communications}\ }\textbf {\bibinfo
  {volume} {159}},\ \bibinfo {pages} {157} (\bibinfo {year}
  {2004})}\BibitemShut {NoStop}%
\bibitem [{\citenamefont {Podest\`{a}}\ \emph
  {et~al.}(2016{\natexlab{a}})\citenamefont {Podest\`{a}}, \citenamefont
  {Gorelenkova}, \citenamefont {Fredrickson}, \citenamefont {Gorelenkov},\ and\
  \citenamefont {White}}]{PodestaNF16}%
  \BibitemOpen
  \bibfield  {author} {\bibinfo {author} {\bibfnamefont {M.}~\bibnamefont
  {Podest\`{a}}}, \bibinfo {author} {\bibfnamefont {M.}~\bibnamefont
  {Gorelenkova}}, \bibinfo {author} {\bibfnamefont {E.~D.}\ \bibnamefont
  {Fredrickson}}, \bibinfo {author} {\bibfnamefont {N.~N.}\ \bibnamefont
  {Gorelenkov}}, \ and\ \bibinfo {author} {\bibfnamefont {R.~B.}\ \bibnamefont
  {White}},\ }\href@noop {} {\bibfield  {journal} {\bibinfo  {journal} {Nuclear
  Fusion}\ }\textbf {\bibinfo {volume} {56}},\ \bibinfo {pages} {112005}
  (\bibinfo {year} {2016}{\natexlab{a}})}\BibitemShut {NoStop}%
\bibitem [{\citenamefont {Podest\`{a}}\ \emph
  {et~al.}(2016{\natexlab{b}})\citenamefont {Podest\`{a}}, \citenamefont
  {Gorelenkova}, \citenamefont {Fredrickson}, \citenamefont {Gorelenkov},\ and\
  \citenamefont {White}}]{PodestaPoP16}%
  \BibitemOpen
  \bibfield  {author} {\bibinfo {author} {\bibfnamefont {M.}~\bibnamefont
  {Podest\`{a}}}, \bibinfo {author} {\bibfnamefont {M.}~\bibnamefont
  {Gorelenkova}}, \bibinfo {author} {\bibfnamefont {E.~D.}\ \bibnamefont
  {Fredrickson}}, \bibinfo {author} {\bibfnamefont {N.~N.}\ \bibnamefont
  {Gorelenkov}}, \ and\ \bibinfo {author} {\bibfnamefont {R.~B.}\ \bibnamefont
  {White}},\ }\href@noop {} {\bibfield  {journal} {\bibinfo  {journal} {Phys.
  Plasmas}\ }\textbf {\bibinfo {volume} {23}},\ \bibinfo {pages} {056106}
  (\bibinfo {year} {2016}{\natexlab{b}})}\BibitemShut {NoStop}%
\bibitem [{\citenamefont {White}\ and\ \citenamefont
  {Chance}(1984)}]{WhitePoP84}%
  \BibitemOpen
  \bibfield  {author} {\bibinfo {author} {\bibfnamefont {R.~B.}\ \bibnamefont
  {White}}\ and\ \bibinfo {author} {\bibfnamefont {M.~S.}\ \bibnamefont
  {Chance}},\ }\href@noop {} {\bibfield  {journal} {\bibinfo  {journal} {Phys.
  Fluids}\ }\textbf {\bibinfo {volume} {27}},\ \bibinfo {pages} {2455}
  (\bibinfo {year} {1984})}\BibitemShut {NoStop}%
\bibitem [{\citenamefont {White}(2011)}]{WhitePPCF11}%
  \BibitemOpen
  \bibfield  {author} {\bibinfo {author} {\bibfnamefont {R.~B.}\ \bibnamefont
  {White}},\ }\href@noop {} {\bibfield  {journal} {\bibinfo  {journal} {Plasma
  Phys. Control. Fusion}\ }\textbf {\bibinfo {volume} {53}},\ \bibinfo {pages}
  {085018} (\bibinfo {year} {2011})}\BibitemShut {NoStop}%
\bibitem [{\citenamefont {White}(2012)}]{WhiteCSNS12}%
  \BibitemOpen
  \bibfield  {author} {\bibinfo {author} {\bibfnamefont {R.~B.}\ \bibnamefont
  {White}},\ }\href@noop {} {\bibfield  {journal} {\bibinfo  {journal} {Commun.
  Nonlinear Sci. Numer. Simulat.}\ }\textbf {\bibinfo {volume} {17}},\ \bibinfo
  {pages} {2200} (\bibinfo {year} {2012})}\BibitemShut {NoStop}%
\bibitem [{\citenamefont {Gorelenkov}\ \emph {et~al.}(2016)\citenamefont
  {Gorelenkov}, \citenamefont {Heidbrink}, \citenamefont {Kramer},
  \citenamefont {Lestz}, \citenamefont {Podest\protect{\`{a}}}, \citenamefont
  {\protect{Van~Zeeland}},\ and\ \citenamefont {White}}]{GorelenkovNF16cgm}%
  \BibitemOpen
  \bibfield  {author} {\bibinfo {author} {\bibfnamefont {N.~N.}\ \bibnamefont
  {Gorelenkov}}, \bibinfo {author} {\bibfnamefont {W.~W.}\ \bibnamefont
  {Heidbrink}}, \bibinfo {author} {\bibfnamefont {G.~J.}\ \bibnamefont
  {Kramer}}, \bibinfo {author} {\bibfnamefont {J.~B.}\ \bibnamefont {Lestz}},
  \bibinfo {author} {\bibfnamefont {M.}~\bibnamefont {Podest\protect{\`{a}}}},
  \bibinfo {author} {\bibfnamefont {M.~A.}\ \bibnamefont
  {\protect{Van~Zeeland}}}, \ and\ \bibinfo {author} {\bibfnamefont {R.~B.}\
  \bibnamefont {White}},\ }\href
  {http://stacks.iop.org/0029-5515/56/i=11/a=112015} {\bibfield  {journal}
  {\bibinfo  {journal} {Nucl. Fusion}\ }\textbf {\bibinfo {volume} {56}},\
  \bibinfo {pages} {112015} (\bibinfo {year} {2016})}\BibitemShut {NoStop}%
\bibitem [{\citenamefont {Gorelenkov}\ \emph
  {et~al.}(1999{\natexlab{b}})\citenamefont {Gorelenkov}, \citenamefont
  {Cheng},\ and\ \citenamefont {Fu}}]{GorelenkovPoP99}%
  \BibitemOpen
  \bibfield  {author} {\bibinfo {author} {\bibfnamefont {N.~N.}\ \bibnamefont
  {Gorelenkov}}, \bibinfo {author} {\bibfnamefont {C.~Z.}\ \bibnamefont
  {Cheng}}, \ and\ \bibinfo {author} {\bibfnamefont {G.~Y.}\ \bibnamefont
  {Fu}},\ }\href@noop {} {\bibfield  {journal} {\bibinfo  {journal} {Phys.
  Plasmas}\ }\textbf {\bibinfo {volume} {6}},\ \bibinfo {pages} {2802}
  (\bibinfo {year} {1999}{\natexlab{b}})}\BibitemShut {NoStop}%
\bibitem [{\citenamefont {Podest\protect{\`{a}}}\ \emph
  {et~al.}(2015)\citenamefont {Podest\protect{\`{a}}}, \citenamefont
  {Gorelenkova}, \citenamefont {Darrow}, \citenamefont {Fredrickson},
  \citenamefont {Gerhardt},\ and\ \citenamefont {White}}]{PodestaNF15}%
  \BibitemOpen
  \bibfield  {author} {\bibinfo {author} {\bibfnamefont {M.}~\bibnamefont
  {Podest\protect{\`{a}}}}, \bibinfo {author} {\bibfnamefont {M.~V.}\
  \bibnamefont {Gorelenkova}}, \bibinfo {author} {\bibfnamefont {D.~S.}\
  \bibnamefont {Darrow}}, \bibinfo {author} {\bibfnamefont {E.~D.}\
  \bibnamefont {Fredrickson}}, \bibinfo {author} {\bibfnamefont {S.~P.}\
  \bibnamefont {Gerhardt}}, \ and\ \bibinfo {author} {\bibfnamefont {R.~B.}\
  \bibnamefont {White}},\ }\href@noop {} {\bibfield  {journal} {\bibinfo
  {journal} {Nucl. Fusion}\ }\textbf {\bibinfo {volume} {55}},\ \bibinfo
  {pages} {053018} (\bibinfo {year} {2015})}\BibitemShut {NoStop}%
\bibitem [{\citenamefont {Cheng}\ \emph {et~al.}(1985)\citenamefont {Cheng},
  \citenamefont {Chen},\ and\ \citenamefont {Chance}}]{ChengAP85}%
  \BibitemOpen
  \bibfield  {author} {\bibinfo {author} {\bibfnamefont {C.~Z.}\ \bibnamefont
  {Cheng}}, \bibinfo {author} {\bibfnamefont {L.}~\bibnamefont {Chen}}, \ and\
  \bibinfo {author} {\bibfnamefont {M.~S.}\ \bibnamefont {Chance}},\
  }\href@noop {} {\bibfield  {journal} {\bibinfo  {journal} {Ann. Phys.}\
  }\textbf {\bibinfo {volume} {161}},\ \bibinfo {pages} {21} (\bibinfo {year}
  {1985})}\BibitemShut {NoStop}%
\bibitem [{\citenamefont {Cheng}(1992)}]{ChengPhR92}%
  \BibitemOpen
  \bibfield  {author} {\bibinfo {author} {\bibfnamefont {C.~Z.}\ \bibnamefont
  {Cheng}},\ }\href@noop {} {\bibfield  {journal} {\bibinfo  {journal} {Phys.
  Reports}\ }\textbf {\bibinfo {volume} {211}},\ \bibinfo {pages} {1} (\bibinfo
  {year} {1992})}\BibitemShut {NoStop}%
\bibitem [{Note1()}]{Note1}%
  \BibitemOpen
  \bibinfo {note} {In NOVA-K, both potential and kinetic energies are defined
  as twice their actual values. This can be seen from comparing equations
  (3.69) and (3.70) of \cite {ChengPhR92} with equations (4.19) and (4.31) of
  \cite {WhiteTH13}. This choice does not introduce changes to the growth
  rate.}\BibitemShut {Stop}%
\bibitem [{\citenamefont {Berk}\ \emph
  {et~al.}(1996{\natexlab{b}})\citenamefont {Berk}, \citenamefont {Breizman},\
  and\ \citenamefont {Pekker}}]{BerkPRL96}%
  \BibitemOpen
  \bibfield  {author} {\bibinfo {author} {\bibfnamefont {H.~L.}\ \bibnamefont
  {Berk}}, \bibinfo {author} {\bibfnamefont {B.~N.}\ \bibnamefont {Breizman}},
  \ and\ \bibinfo {author} {\bibfnamefont {M.~S.}\ \bibnamefont {Pekker}},\
  }\href@noop {} {\bibfield  {journal} {\bibinfo  {journal} {Phys. Rev.
  Letters}\ }\textbf {\bibinfo {volume} {76}},\ \bibinfo {pages} {1256}
  (\bibinfo {year} {1996}{\natexlab{b}})}\BibitemShut {NoStop}%
\bibitem [{\citenamefont {Berk}\ and\ \citenamefont
  {Breizman}(1990{\natexlab{b}})}]{BerkBreizman1990c}%
  \BibitemOpen
  \bibfield  {author} {\bibinfo {author} {\bibfnamefont {H.~L.}\ \bibnamefont
  {Berk}}\ and\ \bibinfo {author} {\bibfnamefont {B.~N.}\ \bibnamefont
  {Breizman}},\ }\href {\doibase 10.1063/1.859406} {\bibfield  {journal}
  {\bibinfo  {journal} {Physics of Fluids B: Plasma Physics}\ }\textbf
  {\bibinfo {volume} {2}},\ \bibinfo {pages} {2246} (\bibinfo {year}
  {1990}{\natexlab{b}})},\ \Eprint
  {http://arxiv.org/abs/http://dx.doi.org/10.1063/1.859406}
  {http://dx.doi.org/10.1063/1.859406} \BibitemShut {NoStop}%
\bibitem [{\citenamefont {White}(2013)}]{WhiteTH13}%
  \BibitemOpen
  \bibfield  {author} {\bibinfo {author} {\bibfnamefont {R.~B.}\ \bibnamefont
  {White}},\ }\href@noop {} {\emph {\bibinfo {title} {The Theory of Toroidally
  Confined Plasmas}}},\ \bibinfo {edition} {3rd}\ ed.\ (\bibinfo  {publisher}
  {Imperial College Press, London, UK},\ \bibinfo {year} {2013})\BibitemShut
  {NoStop}%
\end{thebibliography}

%

\end{document}